\documentclass[final,5p,times,twocolumn]{elsarticle}

\usepackage{amssymb}
\usepackage{amsthm}

\usepackage{lineno}

\journal{International Journal of Multiphase Flow}

\usepackage{amsmath,bm, amssymb}
\usepackage{mathrsfs}
\usepackage[ruled,vlined]{algorithm2e}
\usepackage{graphicx}
\usepackage{caption}
\usepackage{epsfig}
\usepackage{epstopdf}
\usepackage{subcaption}
\usepackage{color}
\usepackage{lmodern}
\usepackage[T1]{fontenc}
\usepackage[utf8]{inputenc}
\usepackage{geometry}
\usepackage{subcaption}
\usepackage[colorlinks=true,linkcolor=blue]{hyperref}%
\usepackage{tabularx} 
\usepackage{booktabs} 
\usepackage{textcomp}

\newcommand{\dd}[1]{\mathrm{d}#1}

\begin{document}

\begin{frontmatter}



\title{Particle resuspension from complex multilayer deposits by laminar flows: statistical analysis and modeling}


\author[inria,ubt]{Hao Liu}
\ead{hao.liu@uni-bayreuth.de}

\author[inria]{Mireille Bossy}
\ead{mireille.bossy@inria.fr}

\author[tud]{Bernhard Vowinckel}
\ead{bernhard.vowinckel@tu-dresden.de}

\author[inria]{Christophe Henry\corref{cor1}}
\cortext[cor1]{Corresponding author}
\ead{christophe.henry@inria.fr}

\affiliation[inria]{organization={Universit\'{e} C\^{o}te d'Azur, Inria, CNRS}, 
            addressline={Calisto team}, 
            city={Sophia Antipolis},
            country={France}}

\affiliation[tud]{organization={Technische Universit\"{a}t Dresden, Department of Hydrosciences},
            postcode={01069},
            city={Dresden},
            country={Germany}}

\affiliation[ubt]{organization={Universit\"{a}t Bayreuth, Department},
            city={Bayreuth},
            country={Germany}}

\begin{abstract}
Particle resuspension refers to the physical process by which solid particles deposited on a surface are, first, detached and, then, entrained away by the action of a fluid flow. In this study, we explore the dynamics of large and heavy spherical particles forming a complex sediment bed which is exposed to a laminar shear flow. For that purpose, we rely on fine-scale simulations based on a fully-resolved flow field around individual particles whose motion is explicitly tracked. Using statistical tools, we characterize several features: (a) the overall bed dynamics (e.g. the average particle velocity as a function of the elevation), (b) the evolution of the top surface of the sediment bed (e.g. distribution of the surface elevation or of the surface slope) and (c) the dynamics of individual particles as they detach from or re-attach to the sediment bed (including the frequency of these events, and the velocity difference / surface angle for each event). These results show that particles detach more frequently around the peaks in the top surface of the sediment bed and that, once detached, they undergo short hops as particles quickly sediment towards the sediment bed. A simple model based on the surface characteristics (including its slope and elevation) is proposed to reproduce the detachment ratio.
\end{abstract}

\begin{graphicalabstract}
\includegraphics{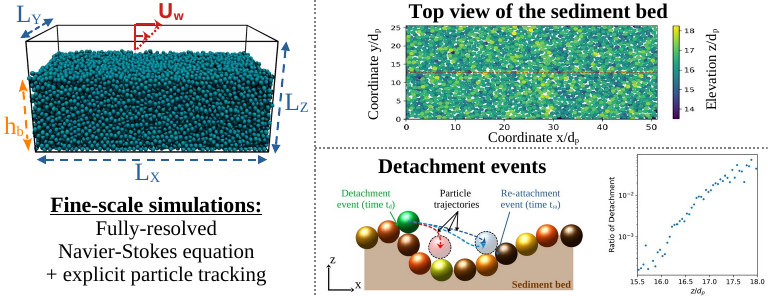}

\end{graphicalabstract}

\begin{highlights}
\item Processing DNS data for a fine-grained statistical analysis of sediment bed dynamics
\item Characterization of the top surface dynamics (bed elevation and local slopes)
\item Characterization of individual detachment/reattachment events: velocity, angle, slope
\item Evidencing short-hops dynamics of the beads, which sediment quickly toward the bed
\item Validation of original simple models reproducing particle velocity and detachment rate
\end{highlights}

\begin{keyword}
Resuspension \sep Multilayer deposit \sep Laminar flow \sep Particle resolved DNS \sep Statistical analysis \sep Model

\end{keyword}

\end{frontmatter}

\section{Introduction}
 \label{sec:intro}

 \subsection{General context}
  \label{sec:intro_context}

When solid particles lying on a surface are exposed to a flow, they can be detached and entrained by the flow: this is the so-called resuspension phenomenon \cite{henry2023particle}. In the last decades, resuspension has received great interest owing to its importance in many environmental and industrial applications. Among the numerous examples that bring out the significance of this topic, one can cite the outbreak of pollutants caused by toxic particles becoming airborne due to wind gusts \cite{brambilla2017adhesion}, the exposure to resuspendable airborne pathogens in indoor environments \cite{rivas2019indoor}, the erosion and dynamics of wind-blown dust/sand \cite{kok2012physics}, the role of sediment entrainment in fluvial environments \cite{pahtz2020physics}, the erosion of riverbanks \cite{lajeunesse2010bed} or the transport of sediments in free-surface flows like rivers \cite{kempe2014relevance}. 

This range of application fields led to various terminologies. For instance, the term `resuspension' is sometimes replaced by `re-aerosolization' in the context of airborne particles \cite{krauter2007reaerosolization}, `remobilization' when dealing with plastics in rivers \cite{liro2020macroplastic} or even `erosion' for sediments in rivers \cite{lajeunesse2010bed}. In the context of aeolian or sediment transport, a specific terminology has even been suggested to distinguish between different modes of transport \cite{kok2012physics}: saltation refers to particles undergoing hops along the surface (i.e. alternating between periods of contact with the surface/bed and periods where they are detached from it), reptation corresponds to particles moving in short hops over the bed (typically less than a centimeter) while creep designates particles remaining in contact with the surface through rolling/sliding motion (including also very slow granular motion associated with sporadic microscopic rearrangements within a bed \cite{pahtz2020physics}). Not to mention the relative importance between the sources of this motion (due to hydrodynamic forces and/or inter-particle collisions \cite{kempe2014relevance}.

For the sake of clarity, we rely here on the notion of particles detached/attached from a surface, independently of their dynamics above the surface as they are detached. In this work, we characterize the detachment and re-attachment of spherical particles from a sediment bed exposed to a laminar shear flow.

 \subsection{Current state-of-the-art and limitations}
  \label{sec:intro_limit}

Despite a century of research that brought forth the key physical mechanisms, a unified model for particle resuspension across the various disciplines and different cases is still lacking \cite{henry2023particle, pahtz2020physics}. This comes partly from the fact that a range of situations are encountered in practical applications, including: (a) particles with sizes ranging from a few micrometers up to a few centimeters, (b) particles displaying various geometries (like spheres, anisotropic particles, complex rough objects), various structural/mechanical properties (e.g. hard or soft, rigid or deformable) or even chemical/biological characteristics, (c) particles forming deposits that are either sparse (i.e. with only a few particles deposited on the surface, leading to so-called monolayer deposits) or dense (possibly giving rise to a large pile, leading to so-called multilayer deposits), and (d) exposed to different types of fluids (e.g. liquid or air) with a range of flows (laminar or turbulent, steady or unsteady).

Nevertheless, significant progress has been made recently thanks to the advances in measurement techniques and in the development of models with increasing complexity to reproduce some of these observations. To be more specific, in the context of monolayer deposits, the motion of isolated particles has been shown to be triggered by three types of events: direct lift-off \cite{shnapp2015comparative, traugott2017experimental}, sliding and rolling \cite{jiang2008characterizing}. This has led to the development of a number of models, based on force-balance, energy-balance or impulse-balance approaches as well as stochastic Lagrangian models \cite{hu2023modeling} that account for the complex dynamics of particles on surfaces (more details can be found in recent reviews \cite{henry2023particle, pahtz2020physics}). More recently, new phenomena have been revealed thanks to the use of high-speed cameras with dense monolayer deposits exposed to unsteady flows \cite{rondeau2021evidence, banari2021evidence}: when the distance between individual particles is comparable to the particle diameter, the detachment of a single particle can lead to multiple detachment events due to a collision-propagation effect (similar to what happens in snooker). The role of inter-particle collision becomes even more predominant when dealing with multilayer deposits, where particles lie on top of one another \cite{henry2023particle}: in such cases, particle resuspension results from a complex interplay between hydrodynamic interactions, inter-particle collisions and inter-particle cohesive forces \cite{lazaridis1998multilayer} that can give rise to various effects (like splashing when one saltating particle leads to multiple detachment due to the energy transferred during its collision with the deposit \cite{beladjine2007collision, oger2008study}). 

While these recent studies pave the way for a unified model, they also highlight the current limitations in our understanding of the phenomena at play and on the limitations of the existing models. In particular, the following issues have been identified (among others): (i) the need for more refined descriptions of surface roughness to account for its effect on particle detachment at small/high velocities \cite{henry2018colloidal, wilson2017influence} and to be more representative of real surfaces that can be contaminated by other objects (like dust in ground surfaces) \cite{rush2018glass, brambilla2020impact}, (ii) a better characterization of the phenomena at play in transient regimes and/or in unsteady flows (e.g. with the occurrence of collision propagation effects) \cite{theron2022influence, cazes2023image, benito2024prediction}, (iii) the development of efficient models that bridge the gap between current formulations used for monolayer or multilayer deposits \cite{henry2023particle, pahtz2020physics}. Despite the complexity of the mechanisms at play in multilayer resuspension (that couple hydrodynamic and collision effects, depend on particle arrangement and deposit structure), recent advances in dense monolayer deposits \cite{rondeau2021evidence, banari2021evidence} provided new ideas to extend previous monolayer models to the case of multilayer cases. 

 \subsection{Objectives of this study}
  \label{sec:intro_aim}

With respect to these recent developments, this study focuses solely on the development of models that bridge the gap between monolayer and multilayer deposits. For that purpose, we rely on the existing Lagrangian stochastic model that has been developed previously in the context of sparse monolayer deposits (see \cite{guingo2008new, henry2014progress, henry2018colloidal, hu2023modeling}) and recently extended to dense monolayer deposits where collision propagation effects arise \cite{banari2021evidence}. This model is retained since it remains compatible with standard approaches used in large-scale CFD (Computational Fluid Dynamics) simulations for engineering purposes (like those encountered for aerosol dispersion or river systems). The stochastic aspect in the model allows to address turbulent flows as detailed in reviews \cite{pope2000turbulent, minier2001pdf, minier2016statistical}. In order to determine how such models have to be adapted to treat the case of multilayer deposits, one requires high-resolution data to identify not only the mechanisms leading to the detachment of particles (hydrodynamic forces or inter-particle collisions) but also to analyze how detachment depends on the local configuration of the bed and how it influences the motion of particles once they are detached. This implies to have access simultaneously to the local fluid velocity around detached particles as well as to the velocity of nearby particles. Despite the recent advances in experimental facilities using high-speed cameras \cite{cazes2023image}, such measurements are not yet available. For that reason, we rely here on the results obtained from fine-scale numerical simulations, where the flow around each individual particle is explicitly resolved (this is the so-called PR-DNS, for particle-resolved direct numerical simulations \cite{tenneti2014particle, uhlmann2023efficient}). Simulations are performed for large particles within a sediment bed exposed to a laminar flow. Due to the high level of information contained in PR-DNS simulations, these simulations are computationally expensive. Drawing on the observations made on the dynamics of individual grains and on the top surface of the bed, we propose simple and computationally efficient model components that reproduces the key features measured all along the three sections. In Sect.~\ref{sec:dr_events}, they are combined together in a single model for detachment rate. 

 \subsection{Layout of the paper}
  \label{sec:intro_layout}

To reach these objectives, we follow a step-by-step approach. The overall dynamics of particles within the sediment bed is first described in Sect.~\ref{sec:part_dyn}, which shows the establishment of a steady-state regime after some time. Then, we characterize the top surface of the sediment bed and its evolution in time in Sect.~\ref{sec:bed_dyn}. Finally, the dynamics of individual particles as they detach or re-attach to the sediment bed is analyzed within the steady-state regime in Sect.~\ref{sec:dr_events} together with a simple model that reproduces the key measurements. 

\section{Dynamics of particles within a sediment bed}
 \label{sec:part_dyn}

The results described in this paper come from the analysis of one of the four numerical simulations that have been described in a previous paper \cite{rettinger2022rheology} and that can be downloaded freely on a \href{https://zenodo.org/records/7857291}{Zenodo repository}. For that reason, we only briefly recall the key information on the system considered in Sect.~\ref{sec:part_dyn:setup}. Section~\ref{sec:part_dyn:method} introduces the methodology used to extract information on the particle dynamics from these simulations while Section~\ref{sec:part_dyn:result} describes the results obtained from this analysis. A model is proposed for the particle velocity in Sect.~\ref{sec:part_dyn:model} to reproduce some of the features measured.

 \subsection{System considered and numerical simulations}
  \label{sec:part_dyn:setup}

\begin{figure}[h]
    \centering
    \includegraphics[width=\columnwidth]{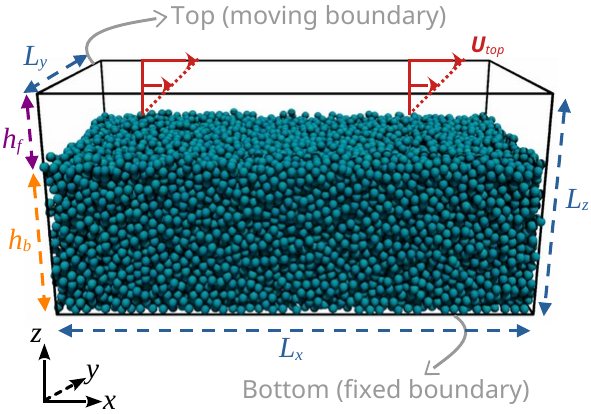}
    \caption{3D-view of the geometry considered here, showing the sediment bed with polydispersed spheres and the initial flow field above the sediment bed obtained by imposing a constant velocity of the top boundary $U_{top}$. Adapted with permission from \cite{rettinger2022rheology}. Copyright 2021, Cambridge University Press. }
    \label{fig:snapshot_sediment}
\end{figure}

We consider the case of a sediment bed exposed to a linear shear flow with a constant shear rate $\dot{\gamma}$, as displayed in Fig.~\ref{fig:snapshot_sediment}. The fluid flow and sediment particles move within a periodic box of size $L_x \times L_y \times L_z = 1024 \times 512 \times 480$. The fluid flow is driven by the motion of the top wall along the $x$-direction at a constant velocity $U_{top} = 0.03$ (in lattice units). The Couette-like flow field obtained is laminar, due to the small Reynolds number based on the channel properties $\textit{Re}_b=0.5\, U_{top} \, h_f /\nu_f\simeq 14$ ($h_f$ being the height of the fluid section and $\nu_f$ the fluid viscosity). A large number of sediment particles is placed initially in the bed (here $N_p=$~26\,112 particles). The particle sizes are generated according to a log-normal distribution, with a mean diameter taken such that $\langle d_p\rangle\simeq20$ (in lattice units) and a variance $\sigma_X^2=0.1$. This distribution is chosen as it is representative of a monodispersed particle suspension, since all particles have a comparable size (the ratio between the largest and smallest particle diameter is 1.15 \cite{rettinger2022rheology}). The particle density $\rho_p$ is taken 1.5 times larger than the fluid density $\rho_f$. 

Numerical simulations are performed by coupling a Lattice-Boltzmann Method (LBM) for the resolution of the fluid flow and a Discrete-Element Method (DEM) for the tracking of sediment particles. The LBM approach describes the evolution of particle distribution functions (PDFs) on a uniform computational grid that fulfills the Navier-Stokes equation for an incompressible Newtonian fluid. This is performed using a lattice with a fixed size $\Delta X = 1$ and a constant time step $\Delta t = 1$ (with the D3Q19 two-relaxation-time model, more details in \cite{rettinger2022efficient}). The DEM approach consists in solving Newton-Euler equations for translational and rotational motion of spheres:
\begin{equation}
\begin{aligned}
    \frac{\dd \bm{X}_{p,i}}{\dd t}  & = \bm{U}_{p,i} \\
    m_{p,i} \frac{\dd \bm{U}_{ p,i}}{\dd t}  &  = \bm{F}_{p,i}^{coll} \, + \, \bm{F}_{p,i}^{hydro} \, + \, \bm{F}_{p,i}^{ext} \\
    I_{p,i} \frac{\dd \bm{\omega}_{p,i}}{\dd t} & = \bm{T}_{p,i}^{coll} \, + \, \bm{T}_{p,i}^{hydro}
\end{aligned}
\end{equation}
where $\bm{X}_{p,i}$ is the position of the center of mass of a particle labeled `i', $\bm{U}_{p,i}$ its velocity, $\bm{\omega}_{p,i}$ its rotation speed, $m_{p,i}$ its mass and $I_{p,i}$ its moment of inertia. These equations take into account various contributions including: inter-particle collision forces $\bm{F}_{p,i}^{coll}$ and torques $\bm{T}_{p,i}^{coll}$ (assuming a soft contact between rigid particles and using a linear spring-dashpot model to handle collisions), hydrodynamic forces $\bm{F}_{p,i}^{hydro}$ and torques $\bm{T}_{p,i}^{hydro}$ as well as the external forces $\bm{F}_{p,i}^{ext}$ (here gravitational and buoyancy forces). These equations are integrated in time using a velocity Verlet scheme with 10 uniform sub-iterations. The coupling between the fluid phase and the particle phase is ensured by the use of a lattice with a grid size roughly 20 times smaller than the mean particle size, meaning that the flow field around each individual particle is explicitly resolved (additional models are used to account for unresolved hydrodynamic forces below the grid size \cite{rettinger2022rheology}). The initial particle bed is obtained by running a precursor simulation where all particles are introduced in the domain, without overlapping, and left to settle down on the surface until all particles come to rest. This leads to a deposit with an average initial height $h_b(0) = 342$ (i.e. with roughly $17$ layers of particles piled up on top of each other).

The data used in the present study contains information on flow fields ($x$, $y$ and $z$ components), averaged along horizontal slabs, as well as on the instantaneous particle quantities (namely the particle radius $r_{p,i}$, the $x$-, $y$-, and $z$-coordinates of the particle's center of mass $\bm{X}_{p,i}$, and its translational velocity $\bm{U}_{p,i}$). Note that we used outputs from the numerical simulation every 1\,000 timesteps and over a total duration of 500\,000 iterations.
Such a high time resolution is required for an unambiguous identification of each detachment/re-attachment events (see Sect.~\ref{sec:dr_events}). In fact, during 1\,000 iterations, particles move over relatively short distances (typically less than $0.1\,d_p$).

 \subsection{Methodology to analyze particle dynamics}
  \label{sec:part_dyn:method}

\begin{figure}[h]
	\centering
	\includegraphics[width=0.8\linewidth]{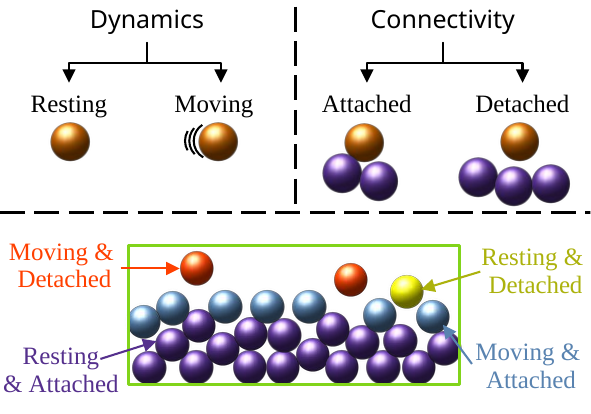}
	\caption{\centering Sketch of the various states in which particles are classified.}
    \label{fig:part_states}
\end{figure}

As displayed in Fig.~\ref{fig:part_states}, particles are classified according to: (a) their state of motion (either resting or moving) and (b) their relation to the sediment bed (either attached to the bed or detached from it). This means that particles are sorted according to four states: (i) resting and attached, (ii) resting and detached, (iii) moving and attached, (iv) moving and detached. By doing so, one can easily distinguish between resuspended and deposited particles.

In practice, the state associated to every particle, along each time snapshot of the PR-DNS simulation, is classified according to these four different classes using an algorithm based on two steps that are detailed below and delineated in the pseudo-code provided in Algorithm~\ref{alg:KDT-G}: 

\begin{algorithm}[h]
\SetAlgoLined
\SetKwComment{Comment}{/*\quad}{*/}
\Comment{Parameters:}
\Comment{\quad $\epsilon_d$: Threshold for distance.}

\KwData{Radii and center coordinates of N particles \(\{...,(r_{p,i},\, x_{p,i},\, y_{p,i},\, z_{p,i}),...\}\).}
\KwResult{"Attached" or "Detached" labeling}
Generate a graph G where each vertice is a particle\;
Reconstruct particles in KDTree\;
\For{i < N}{
    Search for neighbors within spherical domain \((2r_{p,i},\, x_{p,i},\, y_{p,i},\, z_{p,i}) \)\;
    \For{j in neighbors}{
        Compute distance \(d_{ij}\) between \(P_i\) and \(P_j\) \;
        \If{\(d_{ij} \leq (r_{p,i} + r_{p,j}) \, (1 + \epsilon_d)\)}{
            Connect \(P_i\) and \(P_j\) vertices in the graph\;
        }
    }
}
Identify connected sub-graphs in the graph\;
Classify components in G into clusters based on their connections\;
Label particles in the largest cluster as "Attached"\;
Label other particles as "Detached"\;
\caption{Particle Connection Classification Algorithm: KDTtree-Graph}
\label{alg:KDT-G}
\end{algorithm}

\begin{itemize}
    \item The first step is to separate between moving particles and resting ones. To that extent, we introduce a velocity threshold $\epsilon_u$. A particle is considered resting (resp. moving) if its velocity magnitude is smaller (resp. larger) than the shear velocity multiplied by a threshold $\epsilon_u$, i.e. $\|\bm{U}_p\| < U_{top} \, \epsilon_u$ (resp. $\|\bm{U}_p\| \ge U_{top} \, \epsilon_u$). As described in~\ref{app:threshold_vel}, the value of this velocity threshold $\epsilon_u=0.001$ has been carefully chosen with respect to the velocities given by the PR-DNS simulation. In particular, this threshold avoids considering particles deep within the bed as moving ones (these particles might re-arrange at very small velocities but such changes are not of interest in this study where we rather focus on the dynamics close to the top of the sediment bed, where velocities are higher by several orders of magnitude).
    \item The second step consists in sorting particles according to their connectivity with neighboring particles. For that purpose, we rely on a distance threshold $\epsilon_d$ to determine whether two particles are in contact. This threshold $\epsilon_d=0.001$ has been carefully chosen to avoid computational artefacts (see~\ref{app:threshold_dist}). 
    If the separation distance between one particle (labeled~`$i$') and each of its neighbors (labeled~`$j$') exceeds the sum of the two radii multiplied by the threshold, i.e. $d_{ij} > (r_{p,i}+r_{p,j})(1+\epsilon_d)$, the particle is considered as detached. Otherwise, the particle is in contact with at least one other neighbor. Yet, this does not mean that the particle is attached to the sediment bed since it can actually be attached to a group of particles that are moving together above the sediment bed. 
    \newline Hence, an additional piece of information is required to reconstruct the connectivity between particles. To that end, we rely on a KDTree-Graph approach, which allows to handle a large number of particles. KDTree stands as a binary space-partitioning data structure applied for its efficiency in nearest neighbor searches \cite{ram2019revisiting}. Within this framework, a graph is utilized to document the connection among objects. The process involves searching neighboring particles around each target particle within a spherical domain (with a radius equal to the particle diameter) and then computing the distance with these neighbors and identifying which neighbors satisfy the aforementioned distance threshold. This approach ensures efficient distance computations, particularly when dealing with neighboring particles of different sizes—where the distance is calculated only once. Consequently, this optimized strategy significantly reduces the number of distance computations, leading to an algorithm's complexity that scales as $\mathcal{O}(N_p)$ or better ($N_p$ being the number of particles) \cite{ram2019revisiting}. Upon identifying connected particles, the algorithm connects the corresponding vertices in the graph. Once all particles are treated, the resultant graph allows to distinguish between: (1) a substantial connected sub-graph encompassing most components, which represents the bed, (2) small sub-graphs, which correspond to detached particle clusters and (3) disconnected singular components, which are single detached particles. Hence, all particles within the largest sub-graph are labeled as "attached", while the remaining ones are labeled as "detached". 

\end{itemize}

\subsection{Statistics obtained on particle dynamics}
   \label{sec:part_dyn:result}

Thanks to Algorithm~\ref{alg:KDT-G}, one can count the number of particles in each state, every snapshots. In the following we use a dimensionless time $t/t_{\textrm{ref}}$ with $t_{\textrm{ref}}= d_p/U_{top}$ (corresponding roughly to $665$ time steps). The results are displayed in Fig.~\ref{fig:NumStates}, which shows in particular the fraction of moving particles $N_{moving}/N_p$ as a function of the dimensionless time $t/t_{\textrm{ref}}$. Several features can be observed:

\begin{figure}[h]
	\centering
	\includegraphics[width=0.85\linewidth]{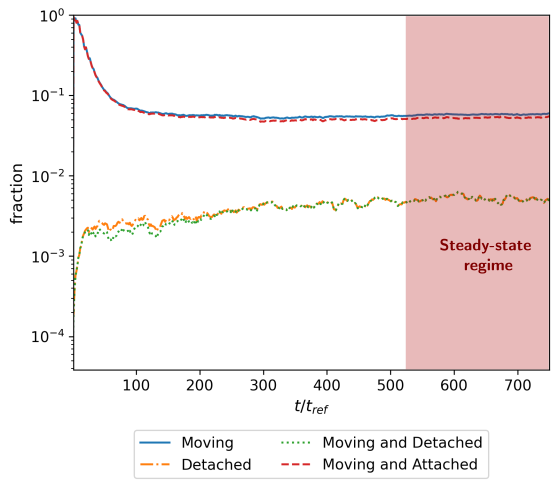}
	\caption{\centering Proportion of particles in each state as a function of the dimensionless time $t/t_{\textrm{ref}}$. }
\label{fig:NumStates}
\end{figure}

\begin{itemize}
    \item The fraction of moving particles quickly rises at the beginning of the simulation, reaching a peak value close to unity before decreasing. This peak in the number of moving particles at the beginning comes from the initialization of the sediment bed. In fact, the initial bed has been obtained by letting particles settle in a quiescent flow until all of them are at rest. Hence, when the shear flow is activated, particles re-organize to a more stable deposit morphology under the action of the flow. Since their velocity becomes higher than the prescribed velocity threshold, we end up having nearly all particles considered as moving in the initial stage. 
    \item After a transition regime ($t/t_{\textrm{ref}} \sim 120$), the number of moving particles becomes nearly constant in time (with a plateau around 4.6\%, corresponding to $\simeq 1\,200$ moving particles). 
    \item Most of these moving particles are actually attached with the sediment bed (the difference between the two being usually smaller than $10\%$). 
    \item There are some particles that are resting and detached in this transition regime. While this case seems counter-intuitive at first, it actually comes from the re-organization of particles deep within the sediment bed: microscopic changes in their position can lead to their detachment for a brief amount of time, but their velocity is so small that they are considered as resting by our current algorithm. Such cases of resting and detached particles are not further studied here since we will focus on the steady-state regime (where they do not occur).
    \item The number of particles that are detached from the sediment bed (orange and green curves) follows a different trend: it quickly rises to reach 0.3\% (i.e. 80-90 particles) in a time $t/t_{\textrm{ref}} \sim 30$, followed by a slow increase until it reaches a statistically-steady regime after $t/t_{\textrm{ref}} \sim 525$ (where close to 0.4\% of the particles are detached, i.e. roughly 100 particles). 
\end{itemize}

In the following, we focus on this steady-state regime and we extract information on particle properties (e.g. velocity, number concentration) averaged over the whole steady-state regime (here for $t/t_{\textrm{ref}}$ between 525 and 750). For that purpose, we simply subdivide the domain into a number of horizontal slabs of size $L_x \times L_y \times 1$ stacked vertically upon each other. Then, within each slab, we compute three quantities: (a) the average fluid volume (also called void fraction or porosity) using the information on the center of mass and the diameter of each particle; (b) the average number of neighbors around each particle (neighbors being defined as other particles with a separation distance smaller than 10\% of the radius, this threshold being taken to include not only particles in contact but those close enough to be affected by short-range forces like lubrication) and (c) the average particle velocity.

\begin{figure}[h]
	\centering
	\includegraphics[width=0.9\linewidth]{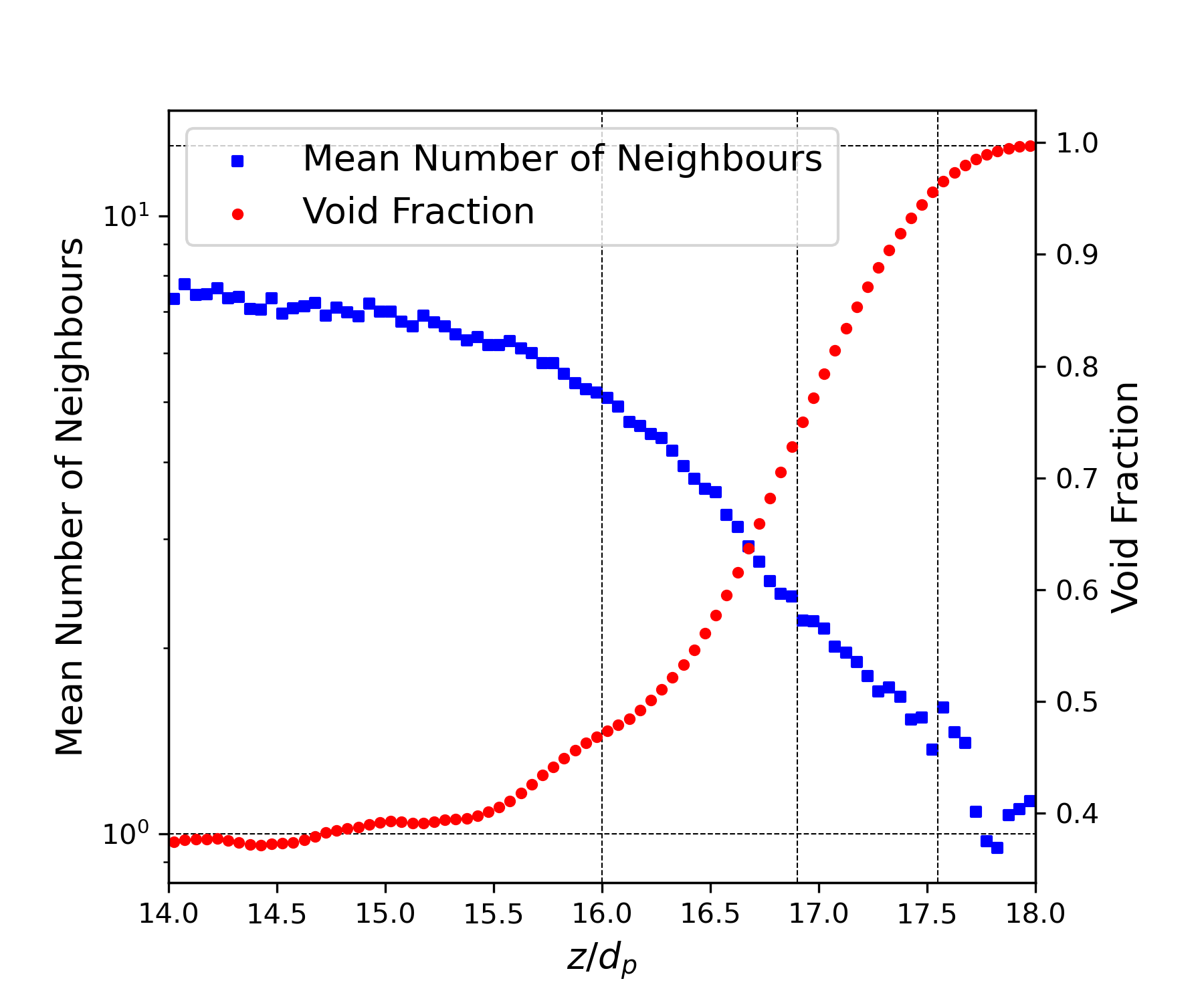}
	\caption{\centering Void fraction (red dots) and mean number of neighbors (blue square) within the sediment bed as a function of the dimensionless elevation $z/d_p$ (note that the average deposit height is $h_b/d_p=17.35$ for the monodispersed case).}
    \label{fig:PartVolFraction}
\end{figure}
Figure~\ref{fig:PartVolFraction} displays the results obtained for the particle void fraction and the number of neighbors. We recover the well-known exponential increase of the void fraction as we go deeper in the sediment bed (see the red curve between $z/d_p \in [16.5,\, 17.5]$), until reaching a constant value around 0.378 in the core of the sediment bed (in agreement with typical values obtained for spheres in a poured random packing) \cite{rettinger2022rheology}. In addition, we measure an exponential increase in the average number of neighbors as we go deeper in the sediment bed (typically for elevations such that $z/d_p \in [16,\, 17.5]$), reaching an average value close to 8 within the core of the sediment bed (a value which is within the range of the various coordination numbers measured in random packings \cite{dullien2012porous}). This exponential increase in the number of neighbors results in an exponential decrease in the void fraction (a feature already confirmed using toy models for adsorption-desorption process \cite{ben1998slow}).

\begin{figure}[h]
	\centering
	\includegraphics[width=0.90\linewidth]{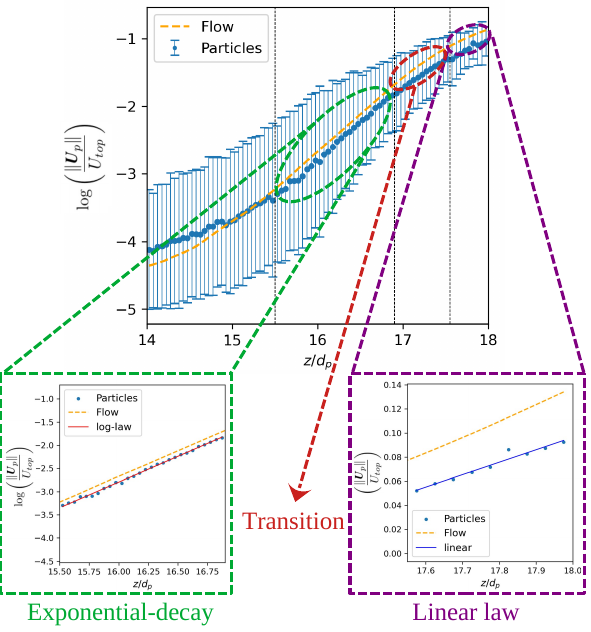}
	\caption{\centering Evolution of the normalized mean fluid velocity and particle velocity ($\|\bm{U}_p\|/U_{top}$) as a function of the dimensionless elevation $z/d_p$. Results are displayed for $z/d_p \in [14,\, 18]$ (this last value being slightly above the height of the top sediment bed).}
\label{fig:PartVel}
\end{figure}
Figure~\ref{fig:PartVel} shows the evolution of the average particle velocity normalized by the top velocity $\|\bm{U}_p\|/U_{top}$ together with the average fluid velocity as a function of the dimensionless elevation $z/d_p$: as seen in \cite{rettinger2022rheology}, the two velocities are closely related. The two insets emphasize four noteworthy regions: (a) the particle velocity profile close to the top of the sediment bed $h_b$ evolves linearly with the vertical height; (b) there is a transition regime when the dimensionless elevation $z/d_p$ is in the range $[17,\, 17.5]$; (c) the particle velocity profile follows an exponential-decay for $z/d_p \in [15.5,\, 17]$; (d) at dimensionless elevations lower than $15.5$, the velocity profile decreases to reach a plateau (where velocities are very small). While the linear increase in the particle velocity near the top of the bed seems to be in line with the hydrodynamic interaction with the laminar shear flow over the sediment bed, the presence of an exponential region can be surprising in a laminar flow even though it only appears over a brief region. The origin of this exponential decay can be related to the exponential increase in the number of neighbors (see the model in Sect.~\ref{sec:part_dyn:model}).

\begin{figure}[h]
	\centering
	\begin{subfigure}{0.9 \linewidth}
	    \includegraphics[width=0.9\linewidth, trim=0cm 0cm 12.5cm 0cm, clip]{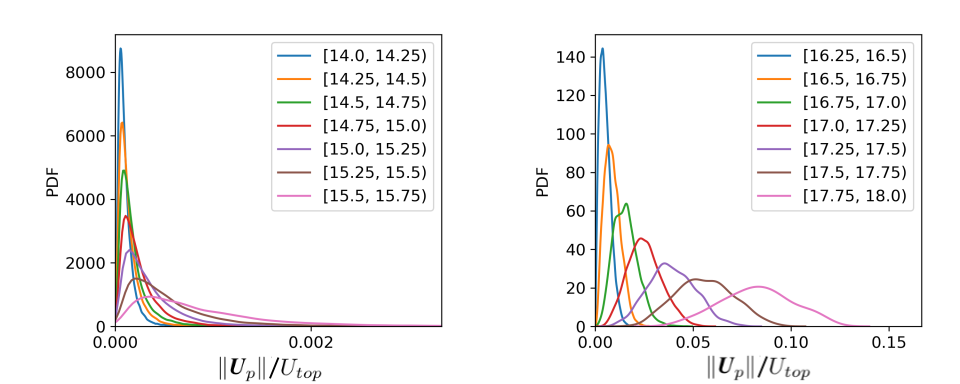}
	    \caption{Distributions of particle velocities $\|\bm{U}_p\|/U_{top}$ for various dimensionless elevations $z/d_p \in [14,\, 15.75]$.}
	\end{subfigure}
	\begin{subfigure}{0.9 \linewidth}
	    \includegraphics[width=0.9\linewidth, trim=12.5cm 0cm 0cm 0cm, clip]{Distribution_Umg.PNG}
	    \caption{Distributions of particle velocities $\|\bm{U}_p\|/U_{top}$ for various dimensionless elevations  $z/d_p \in [16.25,\, 18]$.}
	\end{subfigure}
	\caption{\centering Distributions of the magnitude of the particle velocity $\|\bm{U}_p\|$ sampled at various elevations (horizontal slabs).}
\label{fig:PartVelDistr}
\end{figure}

We have also characterized the distribution of particle velocities within each slab. In fact, knowing only the average particle velocity does not provide information on the extent of their fluctuations. In terms of model development, this can hamper their range of applicability especially if an a-priori law is chosen right from the onset for the distribution of velocities. As shown in Fig.~\ref{fig:PartVelDistr}, the particle velocity distributions resemble normal distributions close to the top of the sediment bed (i.e. $z/d_p \in [17,\, 18]$) while they compare well to log-normal distributions in the region $z/d_p \in [14,\, 15.75]$. This is further confirmed by looking at the evolution of the excess kurtosis and skewness as a function of the elevation (see~\ref{app:PartVel} for more details). 

 \subsection{Model reproducing the particle dynamics}
  \label{sec:part_dyn:model}

In this part, we propose a simple physical model to explain and reproduce the evolution of the particle velocity as a function of the dimensionless elevation $z/d_p$. To be more specific, we focus here on the region close to the top of the deposit bed, where the particle velocities are still relatively large (above an elevation of 15.5). This choice is motivated by the fact that the particle velocities deeper in the bed are at least two orders of magnitude lower than the ones at the top of the bed. As seen from Fig.~\ref{fig:PartVel}, the particle velocity profile displays a linear law above $z/d_p=17.5$ and an exponential decay between $15.5$ and $17$. This can be explained by a model distinguishing two regions:
\begin{itemize}
    \item Particle dynamics driven by hydrodynamic forces: near the top of the bed (at dimensionless elevations higher than $17.5$), particles are mostly set in motion by hydrodynamic interactions with the fluid. These forces include contributions from drag and lift, which arise from the relative velocity between the fluid and a particle. For the sake of simplicity, we disregard the particle inertia in the simple model and thereby assume that the top particles move at the same velocity as the fluid (in reality, due to inertia, particles need time to adapt to the fluid velocity, which would be the terminal velocity if the velocity field was frozen). Since the flow is a simple laminar shear induced by the motion of the top surface at a constant velocity $U_{top} = 0.03$, we choose to model the fluid velocity as a linear shear flow over the average initial height $h_b(0)=342$. This gives:
    \begin{equation}
        U_p(X(t),t) = U_f(X(t),t) = \frac{z_p(t)-h_b(0)}{L_z-h_b(0)} U_{top}.
    \end{equation}
    \item Particle dynamics driven by collisions: deeper in the bed, particles are set in motion by a combination of hydrodynamic forces and inter-particle collisions. In fact, the deeper the particle is, the more likely it is surrounded by neighboring particles and its velocity becomes predominantly driven by energy exchange during collisions. This transition seems to happen here for $z/d_p \in [15.5,\, 17.5]$ (we have retained in the following the highest value to include the transition region in this simple formulation). This is in agreement with the previous study \cite{vowinckel2021incorporating} which showed that contact stresses become predominant over hydrodynamic stresses for a volume fraction of $30\%$ (corresponding here to $z/d_p\sim17$). Drawing on this knowledge, we propose here a layer-by-layer model for particle velocities where the kinetic energy of a layer (`$n$') is given by the exchange of energy with the layer located just above (`$n+1$'). This implies that the kinetic energy in a given layer is directly proportional to the variation in the number of neighbors between the two layers and to the energy dissipated between the two layers (coming from inelastic collisions between individual particles). This gives:
    \begin{equation}
        E_{kin}^{(n)} = E_{kin}^{(n+1)} \, \frac{N_{\rm neighbor}^{(n +1)}}{N_{\rm neighbor}^{(n)}} \, (1-E_{\rm dissip}).
    \end{equation}
\end{itemize}

\begin{figure}[h]
	\centering
	\includegraphics[width=0.95\linewidth]{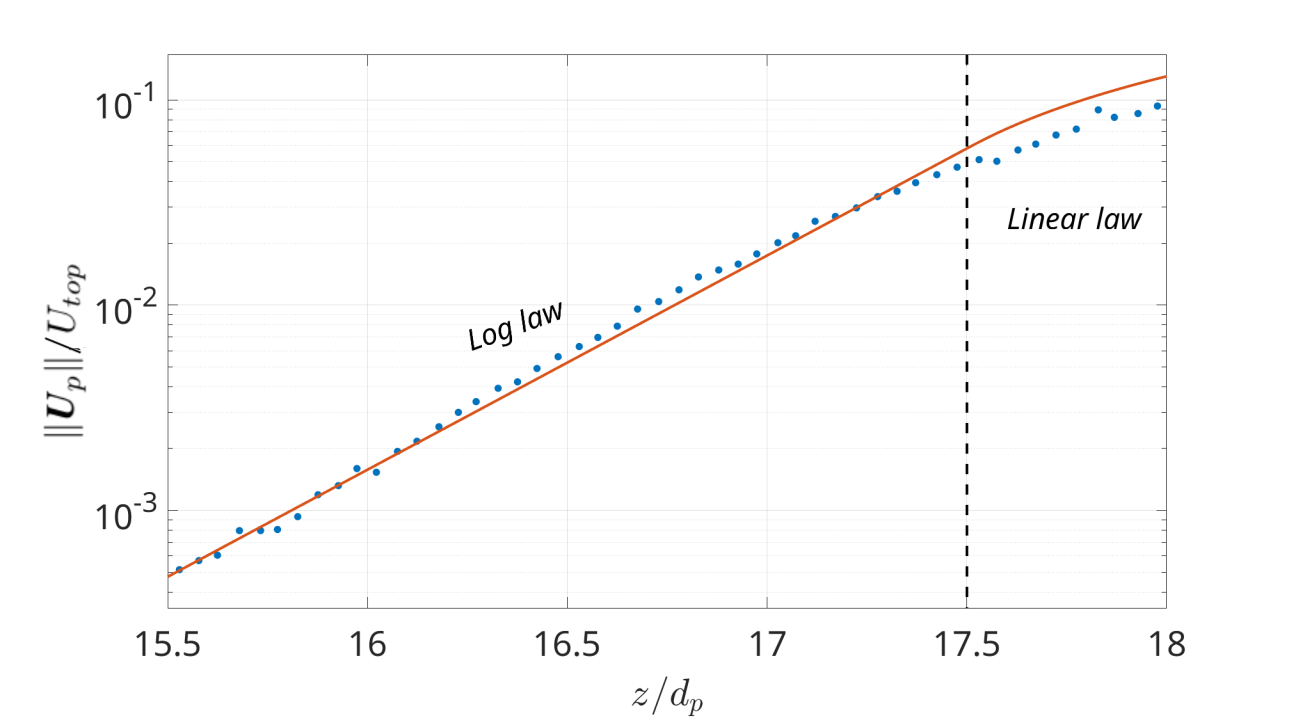}
	\caption{\centering Comparison between the particle velocity measured from PR-DNS simulations (blue dots) and numerical results obtained with the proposed model for particle dynamics (red line).}
    \label{fig:Up_model}
\end{figure}

As displayed in Fig.~\ref{fig:Up_model}, the particle velocity profile can be well reproduced by the model despite its simplicity. However, this good agreement is obtained provided that we use the best exponential fit for the evolution of the mean number of neighbors $N_{\rm neighbor}$ with the elevation $z$ (here estimated with $N_{\rm neighbor} = \text{exp}(-1.2(z/d_p-17.5))$, extracted from Fig.~\ref{fig:PartVolFraction}) and by choosing a value of the energy dissipated between consecutive layers $E_{\rm dissip} = 0.9\%$. The choice of a constant dissipation energy between consecutive layers is motivated by the fact that the particles all have the same size here, meaning that collisions occur between two similar particles regardless of the elevation. In turn, this simple model implies that accurate predictive models for the particle velocity in sediment beds have to properly account for the deposit morphology (here through the mean number of neighbors, but models can also be based on the bed porosity or the void fraction).

\section{Dynamics of the top surface of the sediment bed}
 \label{sec:bed_dyn}

We now turn our attention to the dynamics of the top surface of the sediment bed, whose shape continuously changes with time due to the motion of the particles forming the bed. For that purpose, specific analysis tools have been designed to extract information of interest regarding the particles forming the sediment bed. These tools are presented in Sect.~\ref{sec:bed_dyn:method} before describing the key results obtained in Sect.~\ref{sec:bed_dyn:results}.

 \subsection{Methodology to analyze the top surface of the sediment bed}
  \label{sec:bed_dyn:method}
  
The methodology used to extract information on the local bed configuration at the top of the sediment bed is composed of two steps: the first step is to identify particles located on top of the sediment bed and the second step is to reconstruct the local bed slope. These two steps are described in the following paragraphs. 

The first step consists in identifying all particles that lie on top of the bed. For that purpose, we rely on a straightforward approach based on the information given by the KDTree-Graph method on particles belonging to the bed (see Sect.~\ref{sec:part_dyn:method}): it involves creating a square mesh within the $(xy)$-plane using a predetermined size and then identifying the highest particle within each grid. The outcome of this procedure is obviously sensitive to the grid size $\Delta l$. Taking a value of $\Delta l$ much larger than the particle diameter would lead to the oversight of top particles, resulting in a loss of details in our reconstructed surface. Conversely, using a $\Delta l$ that is much smaller than the particle diameter can lead to erroneous identification of deeper-layer particles as part of the bed surface. To handle this issue, we propose here to use a mesh size slightly smaller than the particle diameter $\Delta l \lesssim d_p$ to ensure an adequate capture of all particles and, subsequently, to employ a Monte Carlo filter to eliminate redundant particles (based on a non-overlapping criteria, where particles are removed if more than 33\% of their area is covered by others when viewed from the top). The procedure is detailed in Algorithm~\ref{alg:reconstruction}.

\begin{algorithm}[h]
\caption{Reconstruction of Bed Surface}
\label{alg:reconstruction}
\SetAlgoLined
\SetKwComment{Comment}{/*\quad}{*/}
\Comment{Parameters:}
\Comment{\quad $\epsilon_{a}$: Threshold area for no-overlap.}

\KwData{Set of particles of the bed}
\KwResult{Set of particles on the bed surface}

Reconstruct $x$- and $y$-coordinates in a 2D KDTree;
Generate a square mesh in the $(xy)$-plane;
Construct a set $\mathcal{P}_h$ of the highest particles within each grid;

\For{$P_i$ in $\mathcal{P}_h$}{
    Search for higher neighbors within the circle $(2r_{p,i},\, x_{p,i},\, y_{p,i})$;
    Generate $N_1$ points randomly inside the circle $(r_{p,i},\, x_{p,i},\, y_{p,i})$;
    Count the number of points inside neighboring circles $N_2$;
    \If{$1 - N_2/N_1$ < $\epsilon_{a}$}{
        Remove $P_i$ from $\mathcal{P}_h$;
    }
}
\end{algorithm}

\begin{figure}[h]
	\centering
	\includegraphics[width=0.75\linewidth]{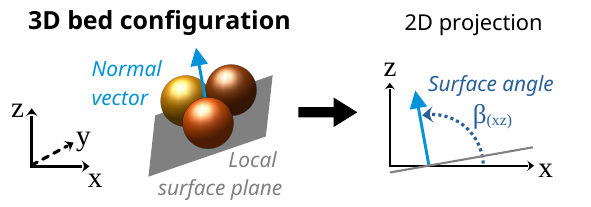}
	\caption{\centering Sketch showing the local surface configuration (left) and the corresponding angles for the surface slope $\beta$ (right).}
\label{fig:sketch_angle_slope}
\end{figure}

The second step is to define the slope of the local surface. For that purpose, we introduce the bed slope angle $\beta$ (see Fig.~\ref{fig:sketch_angle_slope}): it is defined as the angle formed by the vector normal to the plane (i.e. here the plane connecting the center of mass of three neighboring particles) with the $x$-axis. This surface slope $\beta$ thus provides information on the bed orientation: $\beta = \pi /2$ corresponds to a horizontal region of the surface, while $\beta \simeq 0$ (resp. $\beta \simeq \pi$) corresponds to a nearly vertical descending region (resp. nearly vertical ascending region).

 \subsection{Statistics obtained on the dynamics of the top surface}
  \label{sec:bed_dyn:results}
  
\begin{figure}[h]
	\centering
	\includegraphics[width=0.99\linewidth, trim=0cm 0.8cm 0cm 1.0cm, clip]{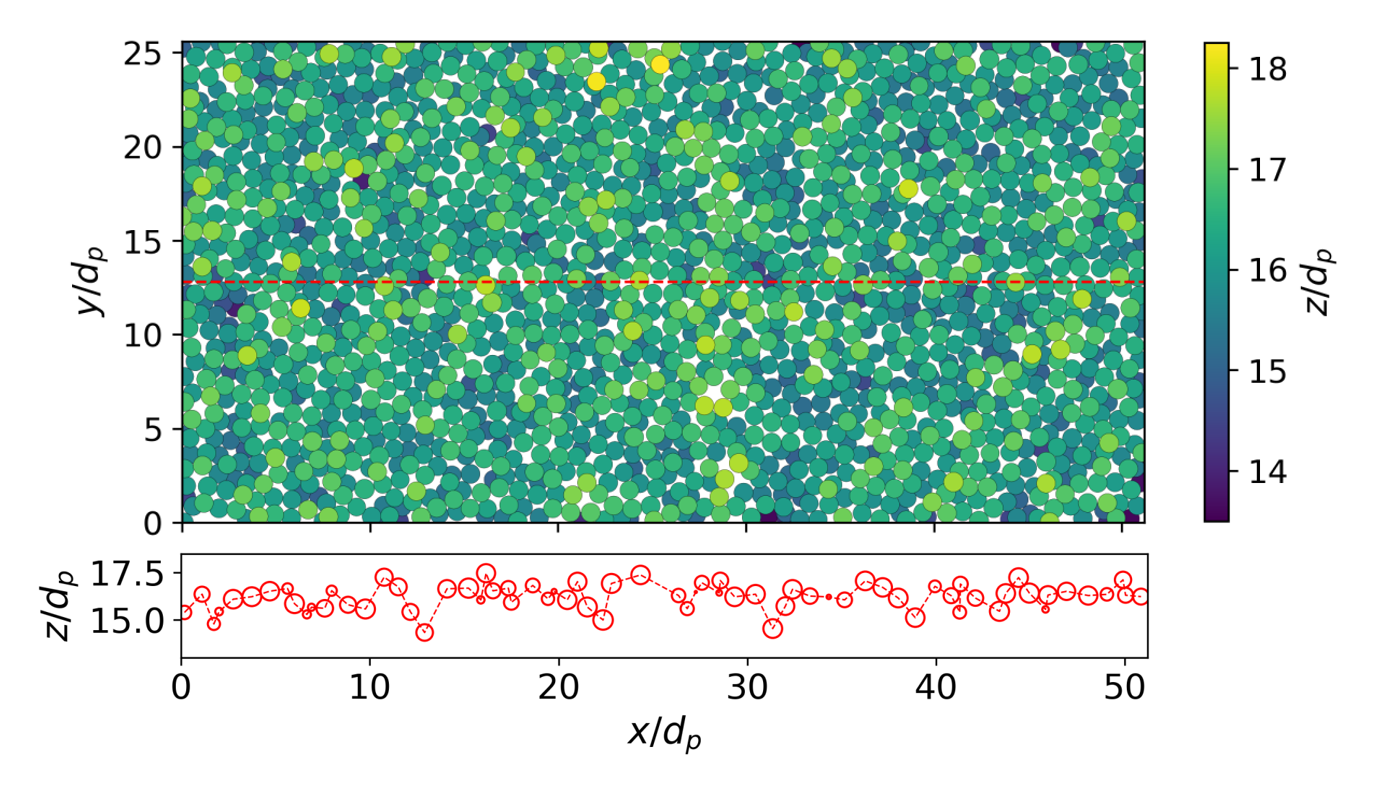}
	\caption{\centering The reconstructed 2D surface at the last time step when $t/t_{ref} = 750$, with particles colored according to the elevation of their center of mass (upper figure) and a 1D profile obtained by a cross section along $y/d_p=12.8$ (lower figure).}
\label{fig:top_surface}
\end{figure}

Figure~\ref{fig:top_surface} illustrates the reconstructed top surface of the bed at the last time step (i.e. here after 500\,000 timesteps which correspond to $t/t_{ref} = 750$): only the top particles are shown in the $(xy)$-plane, and particles are colored according to the elevation of their center of mass. From this 2D map, one can also extract the surface profile along a given line. This is exemplified with the red line at a fixed $y$-coordinate ($y/d_p=12.8$), whose profile is given in the inset below. It shows that the profile along a given line can display sharp fluctuations but that the dimensionless surface height remains between 14.5 and 17.5.

\begin{figure}[h]
	\centering
	\includegraphics[width=0.9\linewidth, trim=0cm 0cm 0cm 1.5cm, clip]{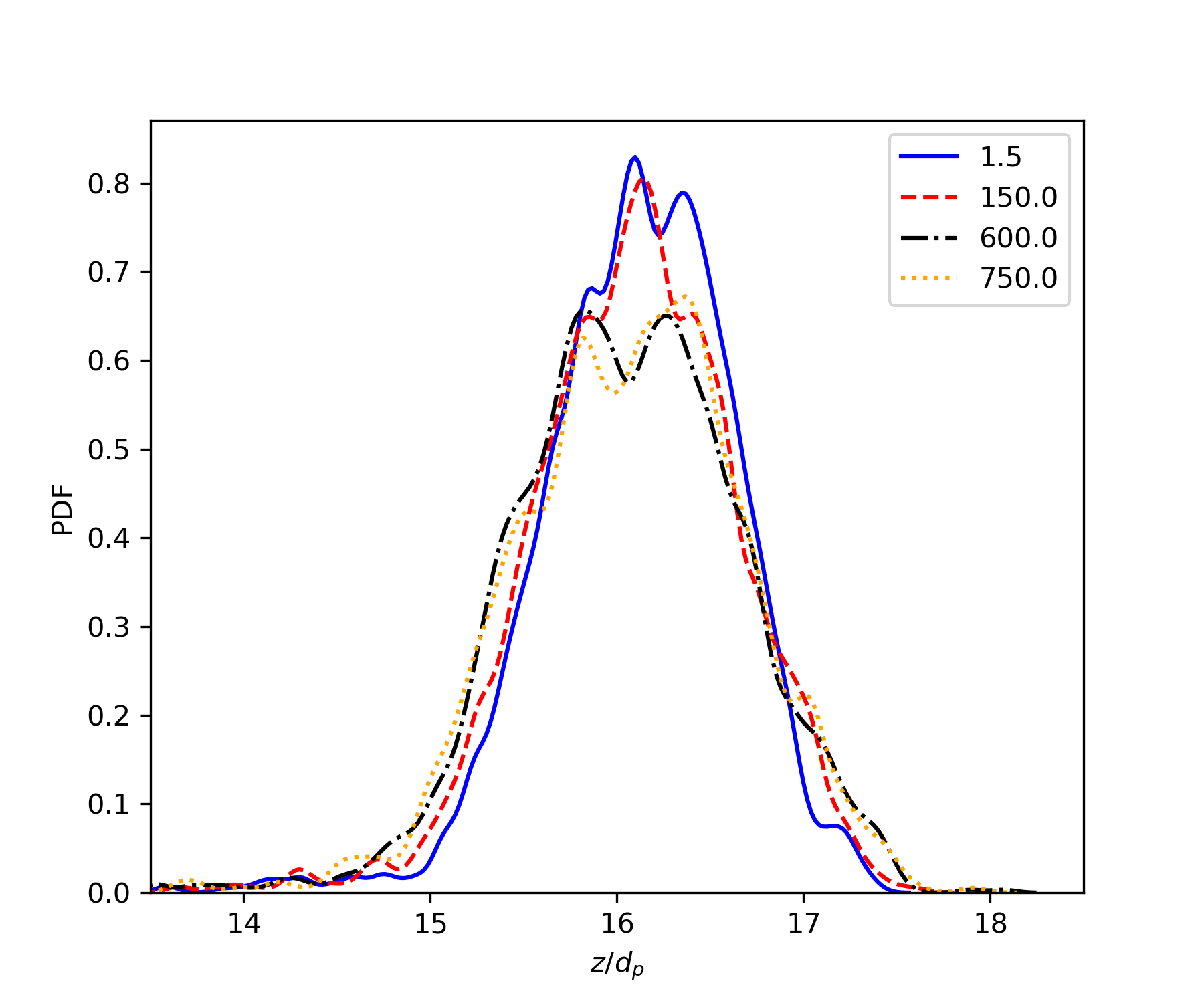}
	\caption{\centering Distribution of the elevation of the top surface at various times (the top view at the last time is visible in Fig.~\ref{fig:top_surface}).}
\label{fig:pdf_surface_height}
\end{figure}

\begin{figure*}[t]
	\centering
	\includegraphics[width=0.75\linewidth]{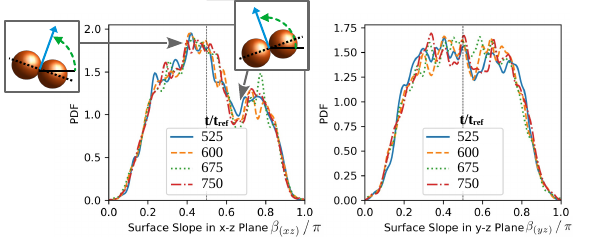}
	\caption{\centering PDF of the surface slope at different times.}
\label{fig:PDF_surf_slope_time}
\end{figure*}

From the reconstructed top surface shown in Fig.~\ref{fig:top_surface}, one can extract statistics on the elevation of this top surface (i.e. here of the elevation of the center of mass of particles forming the top surface). Figure~\ref{fig:pdf_surface_height} displays the distribution of the bed elevation sampled at various times. Several conclusions can be drawn: first, it can be seen that this distribution reaches a steady-state regime after some time (here after $t/t_{ref}=525$, as confirmed by plotting the average, variance, skewness and kurtosis, see~\ref{app:elevation}); second, the steady-state distribution resembles a normal distribution but with two important deviations (one being the presence of two peaks near dimensionless elevations of $15.75$ \& $16.25$ and one being the non-Gaussian tails near the min/max values of the elevation); third, the steady-state distribution is characterized by a negative skewness (see~\ref{app:elevation}), which implies that the surface is composed primarily of valleys and not peaks as explained in \cite{kurth2023systematic}.


The distribution of the surface slope along the $(xz)$- and $(yz)$-planes are displayed in Fig.~\ref{fig:PDF_surf_slope_time} at various times within the statistically steady-state regime (i.e. $t/t_{ref}\ge525$). It appears that the surface slope is symmetric along the $(yz)$-plane around the value $\beta_{(yz)}=\pi/2$. This is consistent with the fact that the shear flow acts along the $(xz)$-plane, meaning that the flow is homogeneous and isotropic along the $y$-direction, which in turns induces a symmetric distribution of surface slopes. Meanwhile, the distribution is asymmetric along the $(xz)$-plane: there is indeed a higher probability to find values of $\beta_{(xz)}\sim 0.4 \pi$ and a lower probability around values close to $0.6 \pi$. This asymmetry originates from the shear flow along the $(xz)$-plane, which induces particle motion along the $x$-direction. However, coming up with a model that explains this trend can only emerge from the analysis of detachment/re-attachment events, whereby particles leave or re-connect to the sediment bed. These events are explored in Sect.~\ref{sec:dr_events}. 

\section{Characterization of detachment and re-attachment events}
 \label{sec:dr_events}

We now turn our attention to the dynamics of each individual particle as they detach or re-attach on the surface. For that purpose, specific analysis tools have been designed to extract information of interest regarding these events and the corresponding particle dynamics. These tools are briefly introduced in Sect.~\ref{sec:dr_events:method} before describing the key results obtained in Sect.~\ref{sec:dr_events:results}. The emerging picture behind this dynamics is provided in Sect.~\ref{sec:dr_events:summary} together with a simple model reproducing the key features (see Sect.~\ref{sec:dr_events:model}).

 \subsection{Methodology to extract statistics on each event}
  \label{sec:dr_events:method}

\begin{figure}[h]
	\centering
	\includegraphics[width=0.8\linewidth]{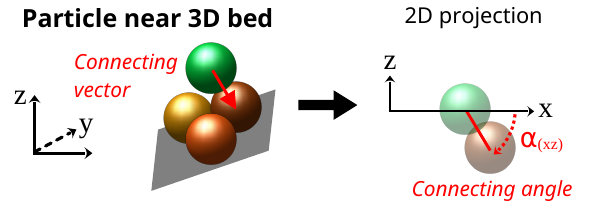}
	\caption{\centering Sketch showing the local surface configuration around a detached particle (left) and the connecting angle $\alpha$ (right).}
\label{fig:sketch_angle_connect}
\end{figure}

Our objective is to characterize the local environment around detached/re-attached particles in order to identify when and where they take place as well as to portray the mechanisms causing such events. For that purpose, we first locate each event in time by detecting as soon as a particle has detached or re-attached to the sediment bed. Then, we rely on the KDTree-Graph method and the connectivity graph (see Sect.~\ref{sec:part_dyn:method}) to pinpoint the closest neighbor in the sediment bed. Based on these two pieces of information, we compute the various statistics of interest for each event, including: (i) the elevation $z$ of the particle center of mass as it detaches/re-attaches, (ii) the local surface slope $\beta$, (iii) the connecting angle $\alpha$, defined as the angle formed by the $x$-axis and the vector connecting the particle to its nearest neighbor within the bed (see Fig.~\ref{fig:sketch_angle_connect}), (iv) the relative velocity with the bed $U_{rel}$, defined as the velocity difference between the particle and its nearest neighbor in the bed and (v) the duration and distance traveled by a particle from its detachment until its re-attachment to the sediment bed.

At this stage, it is important to note that the time and location of each event is not exactly known. In fact, we are using information on particle positions and velocities coming from numerical simulations with outputs every 1\,000 iterations (i.e. $t/t_{ref}=1.5$). This gives rise to some level of uncertainty on the obtained values (which can be lowered using more frequent snapshots in future studies). 

\subsection{Statistics on detachment and re-attachment events}
  \label{sec:dr_events:results}

\begin{figure}[h]
    \centering
    \includegraphics[width=0.8\linewidth, trim=12cm 0cm 0cm 1cm,clip]{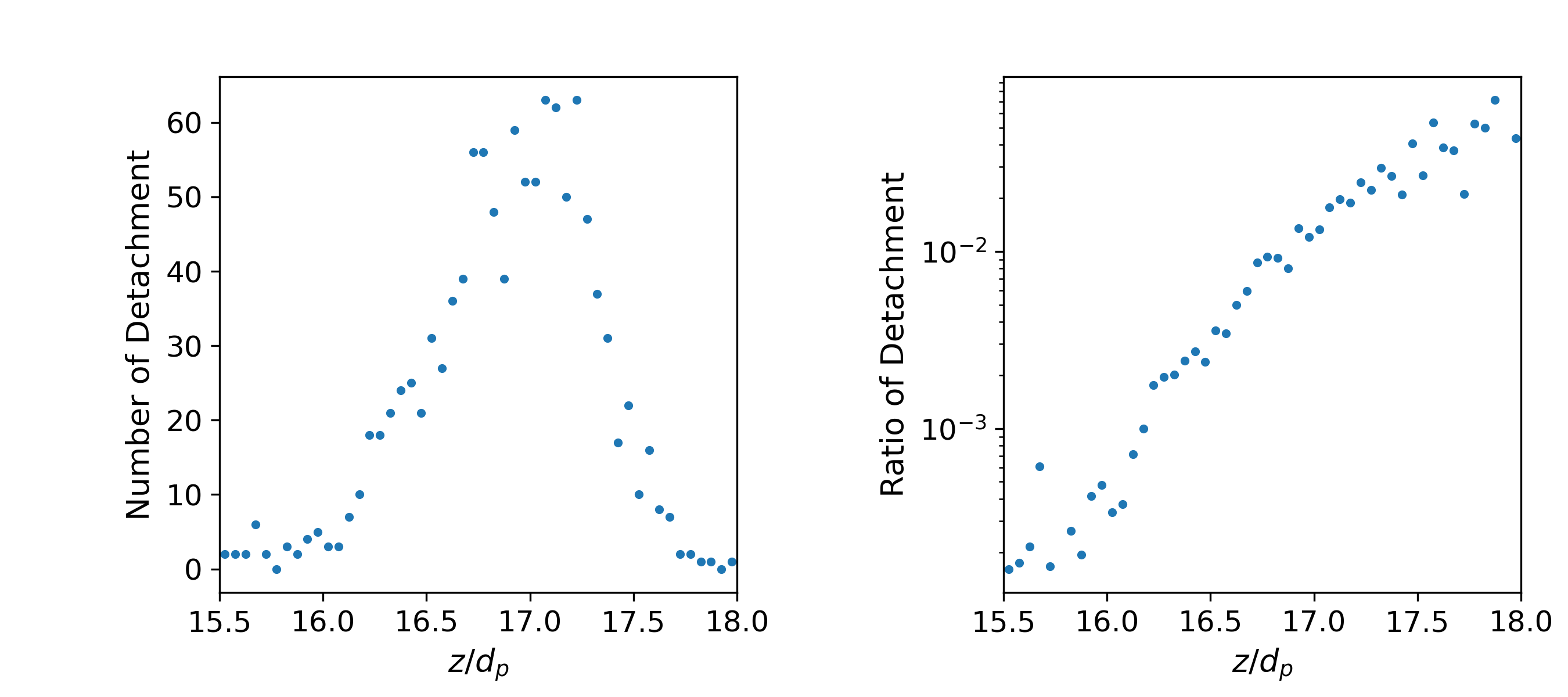}
    \caption{Evolution of the detachment rate (number of detaching events divided by the number of particles within each horizontal slab) as a function of the elevation.}
    \label{fig:DR_event_elevation}
\end{figure}
Firstly, let us start by quantifying how detachment evolves with respect to the elevation.
Figure~\ref{fig:DR_event_elevation} displays how the ratio of detachment (i.e. the number of detaching events divided by the particle number concentration in each horizontal slab) evolves with respect to the elevation $z$. It clearly shows an exponential increase in the detachment rate as a function of the elevation in the range $z/d_p \in [15.5,\, 17.5]$, which corresponds to the region where the top surface of the bed lies (see Sect.~\ref{sec:bed_dyn}). This means that particles located on the highest peak of the top surface detach more easily than the ones located in the deepest valleys of the bed surface. As will be seen later in Sect.~\ref{sec:dr_events:model}, this is related to the fact that particles located at higher elevations are exposed to higher fluid velocities.

\begin{figure}[h]
	\centering
	\includegraphics[width=0.9\linewidth, trim = 3cm 0cm 23cm 1cm, clip]{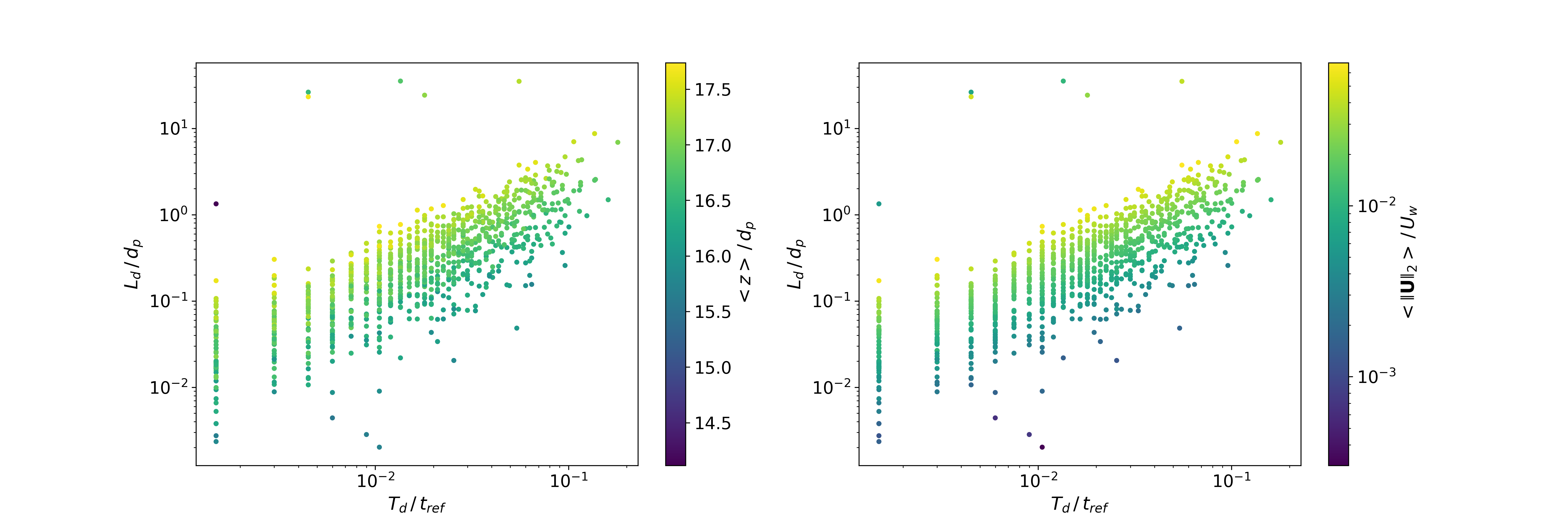}
	\caption{\centering Scatter plot of the duration and distance traveled by particles during a hop-like motion (i.e. between a detachment event and the subsequent re-attachment to the surface). Each point is colored according to the dimensionless elevation of the center of mass of the detached particle.}
\label{fig:DR_event_time_dist}
\end{figure}

Secondly, Fig.~\ref{fig:DR_event_surfslope} displays the distribution of surface slope $\beta_{(xz)}$ conditioned on detachment events (blue curve) and on re-attachment events (red curve). When comparing these distributions to the one obtained for the top surface of the sediment bed (black curve, obtained by averaging over time the samples shown in Fig.~\ref{fig:PDF_surf_slope_time} within the steady-state regime), two main conclusions can be drawn: 
\begin{figure}[h]
	\centering
	\includegraphics[width=0.8\linewidth]{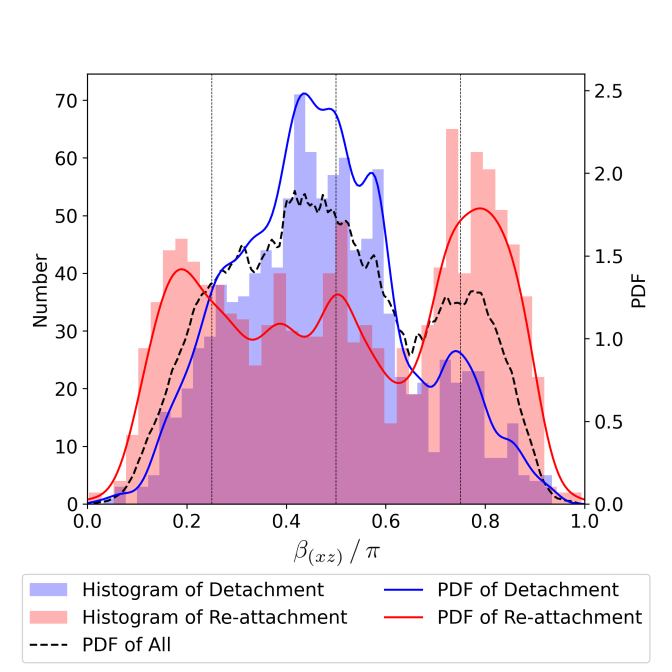}
	\caption{\centering PDF of the surface slope: comparison of the distributions obtained for the top surface of the bed (black line) and the ones conditioned solely on detachment events (blue) or re-attachment events (red).}
\label{fig:DR_event_surfslope}
\end{figure}
\begin{enumerate}[i.]
    \item Particles tend to detach more frequently in regions close to a local extremum, i.e. with a surface slope $\beta_{(xz)}\sim\pi/2$ that is representative of a horizontal (flat) surface. Correlating this observation to the exponential increase in the detachment ratio shown in Fig.~\ref{fig:DR_event_elevation}, it appears that particles detach predominantly near peaks in the top surface of the sediment bed.
    \item Particles re-attach more frequently in rather steep descending/ascending regions, i.e. here with surface slopes typically around $\beta_{(xz)}\sim\pi/2\pm\pi/4$ that are representative of regions tilted by $\pm 45$\textdegree. As will appear later, this has to do with the dynamics of the particles once they are detached from the sediment bed: due to the large particle size and to the large relative density $\rho_p/\rho_f = 1.5$ (which gives a Stokes number $St=\rho_p\,d_p^2\dot{\gamma}/\nu_f$ around 0.85 \cite{rettinger2022rheology}), particles sediment rather quickly (see Fig.~\ref{fig:DR_event_time_dist}). Hence, upon re-attaching, these particles sample preferentially descending/ascending regions, contrary to light saltating particles that display long hop-like motion (which would normally lead to their re-attachment in any region of the surface).
\end{enumerate}

Thirdly, the distribution of connecting angles is shown in Fig.~\ref{fig:DR_event_connectangle}. It compares the values obtained solely for detaching particles (red curve) or re-attaching ones (orange curve) to the values obtained for a fictitious particle (which has been used to probe the top surface of the sediment bed with a regular spacing along $x$- and $y$-directions and forcing the contact with the bed before computing the connecting angle). The results for the fictitious particle in contact with the top surface show that the distribution is close to a normal distribution, with the most frequent value around $\alpha_{(xz)}\sim-\pi/2$ (which corresponds to the case where the fictitious particle is lying on top of another one from the bed). When comparing it to detachment/re-attachment events, various observations can be made:
\begin{figure}[h]
	\centering
	\includegraphics[width=0.95\linewidth]{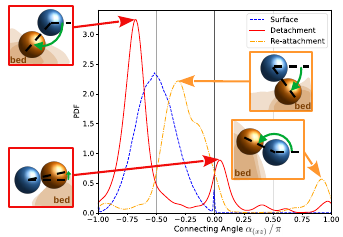}
	\caption{\centering PDF of the connecting angle: comparison of the distributions obtained by probing the top surface with a fictitious particle (blue) and the ones conditioned for detachment (red) and re-attachment (orange) events.}
\label{fig:DR_event_connectangle}
\end{figure}

\begin{itemize}
    \item The distribution obtained for detachment events differs significantly. In fact, detachment occurs preferably around $\alpha_{(xz)}\sim-3\pi/4$, which corresponds to detaching particles located in the downstream region with a 45 degree angle from the nearest particle in the bed (see Fig.~\ref{fig:DR_event_connectangle}). In addition, detachment occurs sometimes with a connecting angle $\alpha_{(xz)}\gtrsim0$, i.e. for detaching particles located on the same plane but upstream of the nearest particle in the bed. Such situations are more likely to occur in the regions near a local peak in the top surface of the sediment bed (in line with the previous observation on surface slopes displayed in Fig.~\ref{fig:DR_event_surfslope}). 
    \item The distribution for re-attachment events is again very different, since particles re-attach most frequently with a connecting angle $\alpha_{(xz)}\sim-\pi/4$, which occurs when the re-attaching particle is located in the upstream region with a 45 degree angle from the nearest particle in the bed (see Fig.~\ref{fig:DR_event_connectangle}). Again, another situation where re-attachment can occur, although less frequently, is when the connecting angle $\alpha_{(xz)}\lesssim1$, i.e. for re-attaching particles located downstream of the bed. This second situation can be surprising at first sight, but it actually coincide with cases where a detached particle is sedimenting behind a peak in the top surface. This peak can move faster than the detached particle in the streamwise direction (e.g. due to shielding effects on the detached particle in the wake of the peak).
\end{itemize}

\begin{figure}[h]
	\centering
	\includegraphics[width=0.85\linewidth]{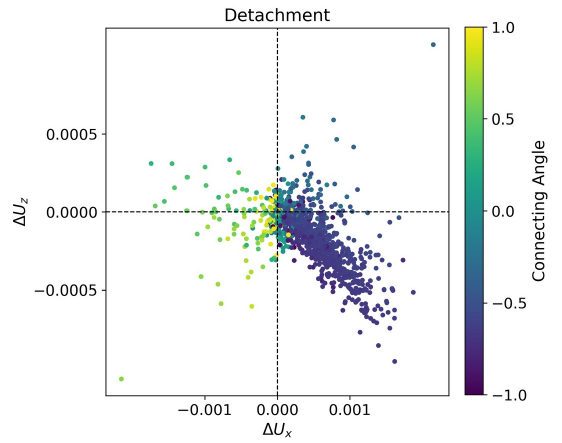}
	\includegraphics[width=0.85\linewidth]{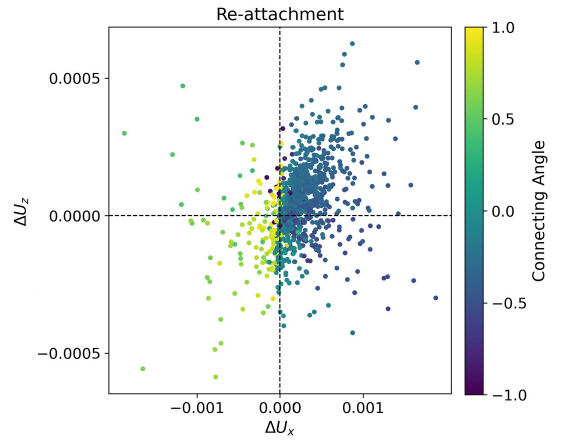}
	\caption{\centering Relative velocities between the detached/re-attached particles and the sediment bed when such events happen. Each point is colored according to the connecting angle.}
\label{fig:DR_event_velocity}
\end{figure}

Fourthly, the relative velocity between the sediment bed and the detached particle along $x$ and $z$ components are shown in Fig.~\ref{fig:DR_event_velocity} for each detachment event (left) and re-attachment event (right). These results clearly show that the vast majority of detachment events are associated with a positive relative velocity along the $x$-direction (meaning that the detached particle moves downstream faster than the sediment bed) but a rather negative relative velocity along the $z$ direction (meaning that detached particles move downward, probably due to sedimentation). Meanwhile, re-attachment events also occur preferably for particles moving faster than the sediment bed along the $x$-direction. Yet, the relative velocity along the $z$-direction can either be negative (corresponding to particles re-attaching through simple settling) or positive (implying that particles are moving upward faster than the nearest neighbor in the sediment bed before re-attaching). In agreement with the previous observation on connecting angles, the relative velocity along the $x$-direction is negative during re-attachment when the connecting angle is close to 1. This confirms the fact that such situations arise when a particle moves in the wake of a peak in the sediment bed which happens to move faster downstream.

Fifthly, the distance traveled by a particle between its detachment point and its re-attachment point to the sediment bed is displayed in Fig.~\ref{fig:DR_event_time_dist} as a function of the duration of the detachment. This plot shows that this hop-like distance is usually smaller than a few particle diameters and is correlated to the time spent detached from the sediment bed. These short distances/times are related to the fact that the particles do settle relatively fast compared to the velocity of the laminar shear flow. Hence, we obtain short hops similar to reptation in aeolian sciences \cite{duran2011aeolian} instead of long hop-like motion characteristic of saltating particles \cite{ho2014aeolian}.

\subsection{Emerging picture for the dynamics}
  \label{sec:dr_events:summary}

\begin{figure}[h]
	\centering
	\includegraphics[width=0.8\linewidth]{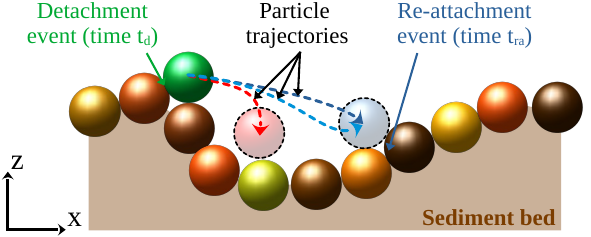}
	\caption{\centering Sketch of the trajectory of a particle as it detaches from the sediment bed (at time $t_{d}$ and later re-attaches to it at time $t_{ra}$. Note that, for the sake of clarity, the rest of the bed is not altered as the particle undergoes such a hop-like motion (this seldom occurs in reality since particles in the bed move through creeping motion, leading to continuous re-arrangements of the deposit morphology due to inter-particle collisions).}
\label{fig:sketch_detach_reattach}
\end{figure}

Combining all these observations together, it appears that particles detach preferably around peaks on top of the sediment bed, while they re-attach in the valleys. The motion of detached particles is driven by the action of the fluid (which drives them in the streamwise direction) as well as by gravitational settling (which drives them toward the bottom regions) and by contact forces (including lubrication forces that tend to reduce the relative velocity even when there is no contact yet). This behavior is typical of the reptation motion in aeolian sciences \cite{kok2012physics}, which is governed by gravity and contact forces. Meanwhile, particles detach preferably near the highest elevations (peaks), where the fluid velocities encountered are the strongest. At this stage, one should note that there is no way to know if detachment is triggered by hydrodynamic forces only, collision or a combination of both. This is due to the time resolution used here which is not high enough to have access to the exact chain of events that can trigger detachment/re-attachment, especially since many particles forming the top surface of the sediment bed are actually moving in a creeping motion (i.e. by rolling/sliding on each other while remaining in contact). This emerging picture is summarized in Fig.~\ref{fig:sketch_detach_reattach}.

Drawing back on the distribution of surface slope displayed in Fig.~\ref{fig:PDF_surf_slope_time}, this picture of particle detachment/re-attachment explains the second peak around $\beta_{(xz)}\sim0.8$ but it fails to account for the lower probability to have surface slope $\beta_{(xz)}\sim0.65$. In fact, this comes from the dynamics of particles forming the top surface, which are continuously moving due to the hydrodynamic force exerted by the fluid flow, to the collisions between particles or to the combination of both. This dynamics tend to push particles in the upward regions towards nearby peaks where they have a high probability to detach. 

\subsection{Model reproducing detachment rates}
  \label{sec:dr_events:model}

In this section, we suggest a simple model to reproduce some of the observations made on the detachment/re-attachment events and on the characteristics of the top surface of the sediment bed. Drawing on existing theories (see reviews in \cite{henry2023particle, henry2014progress, ziskind2006particle}), we speculate that detachment occurs either due to the action of the fluid flow, due to collisions with other particles or a combination of both. Yet, we cannot distinguish easily between the two here due to the time resolution of the simulation output (here with snapshots of particle positions and velocities every 1~000 iterations). Hence, we estimate the particle velocity without taking into consideration the precise source of its motion (this is left out for a future model). 

Following the observations made previously, we propose a simple model that distinguishes two cases (see also Fig.~\ref{fig:DR_model_sketch}):
\begin{figure}[h]
	\centering
	\includegraphics[width=0.99\linewidth]{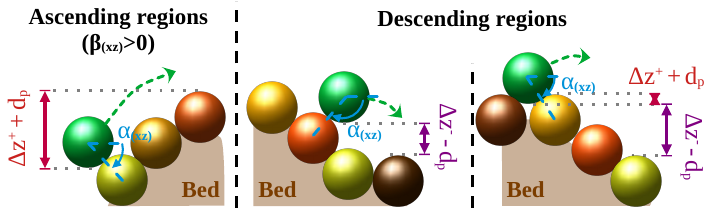}
	\caption{\centering Sketch of the model used for particle detachment in both upward and downward regions of the top surface of the sediment bed.}
\label{fig:DR_model_sketch}
\end{figure}
\begin{itemize}
    \item Particles in ascending regions of the top surface detach if they manage to go beyond the next surface peak along the streamwise direction. In terms of energetic considerations, this amounts to say that the kinetic energy (based on the relative velocity between the detaching particle and the nearby bed) is high enough to overcome the potential energy needed to go above the next surface peak (see Fig.~\ref{fig:DR_model_sketch}): 
    \begin{equation}
        E_{kin} \ge E_{pot}.
    \end{equation}
    This can be re-written in terms of the particle properties and the elevation difference $\Delta z^+$ between the center of the detaching particle and the center of the particle at the next peak of the sediment bed:
    \begin{equation}
        \label{eq:model:cond_ascend}
        \frac{1}{2} \, \rho_p \, \frac{4}{3} \, \pi \, r_p^3 \, (\Delta U_{p,z})^2 \ge \frac{4}{3} \, \pi \, r_p^3 \, \rho_p \, g \, \Delta z^+.
    \end{equation}
    Assuming that the whole bed is moving at the same velocity during the detachment event, this gives a condition for the relative velocity:
    \begin{equation}
        \Delta U_{p,z} \ge \sqrt{2 \, g \, \Delta z^+}.
    \end{equation}
   
    \item Particles in descending regions of the top surface detach if one of the two following conditions is fulfilled. First, if the connecting angle with the nearest particle in the bed is negative, nothing prevents the particle from detaching if its relative velocity along the $x$-direction is positive (i.e. $\Delta U_{p,x} >0$) and if the next valley is lower than the current particle elevation (i.e. $\Delta z^->d_p$, where $\Delta z^-$ is the center-to-center distance between the detaching particle and the particle at the next valley). Second, if the connecting angle is positive, this means that the detaching particle has to go slightly upward to overcome the contact with the nearest particle (see Fig.~\ref{fig:DR_model_sketch}). This implies that the same conditions apply, except for the relative velocity which has to satisfy the following condition (similarly to Eq.~\eqref{eq:model:cond_ascend} for ascending regions):
    \begin{equation}
        \label{eq:model:cond_descend}
        \Delta U_{p,z} \ge \sqrt{2 \, g \, d_p \, \text{sin}\alpha}.
    \end{equation}
    
\end{itemize}

This model requires a-priori information on the relative velocity between the detaching particle and the nearest particle in the bed $\Delta \bm{U}_p$, on the surface slope $\beta$, on the connecting angle $\alpha$ and on the elevation difference with the nearest peak $\Delta z^+$ or the nearest valley $\Delta z^-$ along the streamwise direction. For the sake of simplicity, we have randomly generated these values here using the measurements made before. This means that we generate the velocity of the detaching particle and its nearest neighbor independently of each other following the model for the particle velocity (see Sect.~\ref{sec:part_dyn:model}). The surface slope is generated assuming a continuous distribution in the range $[0.2\pi,\, 0.8\pi]$ (following Fig.~\ref{fig:sketch_angle_connect}) while the connecting angle is drawn within a normal distribution with zero mean and a standard deviation of $2\pi/3$ (following Fig.~\ref{fig:DR_event_connectangle}). Finally, the elevation of the next peak (resp. next valley) is randomly chosen within a normal distribution centered around $z/d_p=16.25$ and with a standard deviation equal to $0.42$ (following Fig.~\ref{fig:top_surface}) and then retaining only the values above (resp. below) the elevation of the detaching particle.

\begin{figure}[h]
	\centering
	\includegraphics[width=0.98\linewidth]{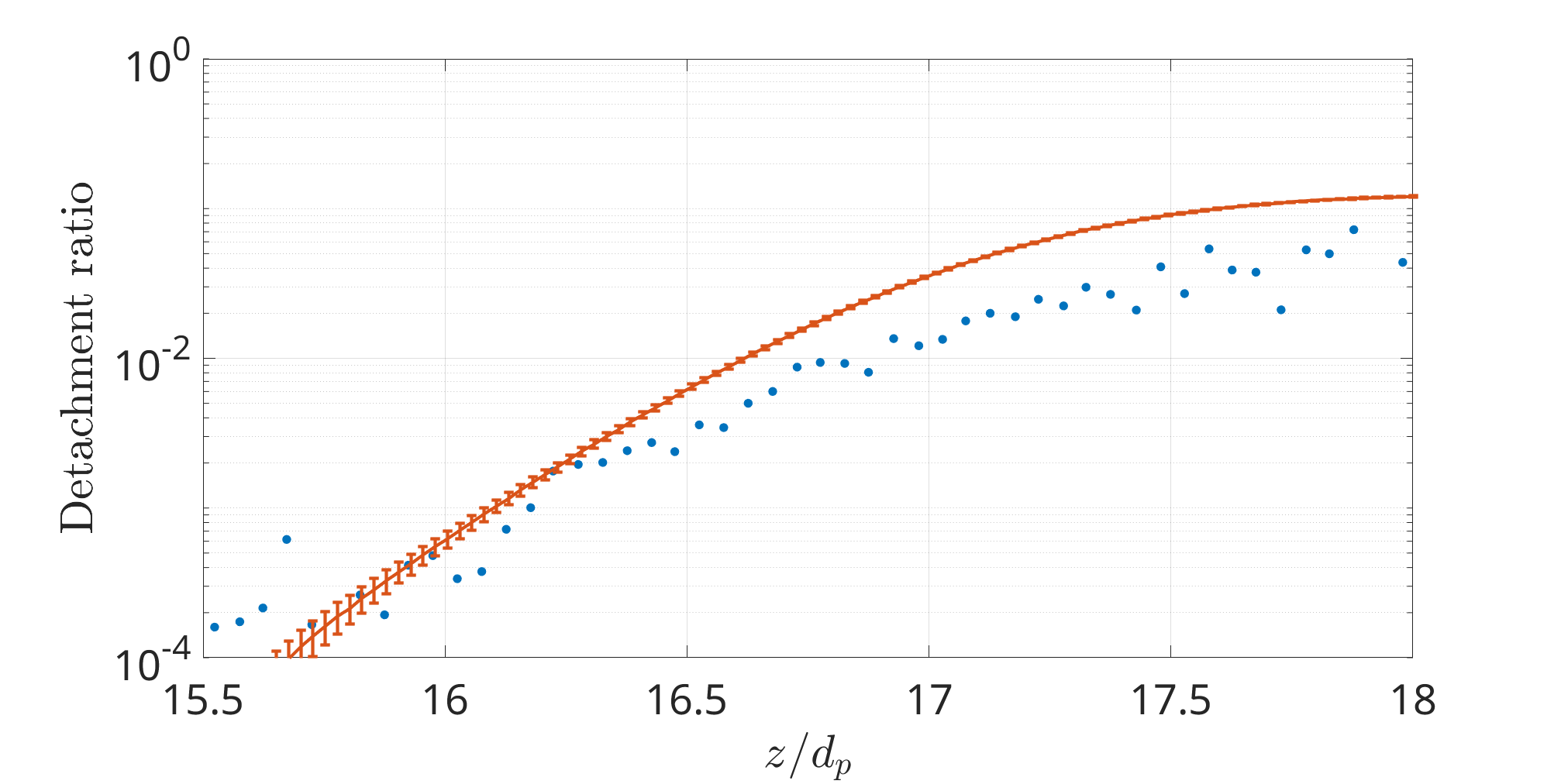}
	\caption{\centering Evolution of the detachment ratio with the dimensionless elevation $z/d_p$: comparison between the model predictions (red line, mean value obtained over 1~000 independent realizations with error bars) and the extracted data (blue dots).}
\label{fig:detach_model_comp}
\end{figure}

The results obtained with this model are displayed in Fig.~\ref{fig:detach_model_comp}, which compares the detachment ratio measured from the simulations to the one obtained from the model (with the average value and standard deviation over 1~000 independent realizations). The good agreement obtained confirms the need to develop accurate models that should account for both the particle velocity and the bed morphology. More precisely, this implies that existing Lagrangian stochastic models for monolayer deposits should include information on the sediment bed, including in particular statistics on the surface slope, on the surface elevation and on the connecting angle with attached particles. Such developments will be undergone in future studies, once the initiation of motion has been more clearly identified. 

\section{Conclusion}
 \label{sec:conclusion}

In this work, we have explored the dynamics of large and heavy spherical particles forming a complex sediment bed exposed to a laminar shear flow using fine-scale simulations (based on a fully-resolved flow field around individual particles whose motion is explicitly tracked). Using statistical tools together with KDTree-Graph approaches, we have characterized several aspects. First, the overall bed dynamics revealed the existence of a statistically-steady regime for $t/t_{ref}>525$, in which the particle velocity profile displays a linear law at the top of the sediment bed and an exponential decay deeper within the bed (typically between $z/d_p \in [15.5,\, 17]$). Second, the elevation of the bed top surface was shown to display a distribution close to a normal one while the surface slope is not homogeneous in the $(xz)$-plane. This dynamics of the top surface has been related to the detachment and re-attachment events occurring constantly, especially for particles located close to the peak regions of the top surface and provided that the relative velocity is positive along the $x$-direction. This analysis also revealed that detached particle only display short hop-like motion since they quickly settle once detached. Based on these observations, a simple model based on the surface characteristics (including its slope and elevation) has been proposed to reproduce the detachment ratio measured. 

While this model provides a relatively good agreement with the measurements made from fine-scale simulations, it is only a first step towards more advanced models based on stochastic Lagrangian formulations that can cover both monolayer and multilayer cases. In fact, there are still uncertainties on what are exactly the causes behind the detachment of particles (hydrodynamic interactions or inter-particle collisions). Investigating such issues will be done in the near future, using new simulations with a much higher frequency on the output results so as to identify each event precisely. In addition, information on the rotational velocity of the particle will allow us to distinguish between sliding and rolling particles, while developing a more complete model for particle collisions that includes both translational and rotational motion (in the spirit of \cite{cross2019oblique}). Finally, this model neglects correlations between the velocities of neighboring particles: this will be investigated in future works, which will also include turbulent flows, where such correlations can play a significant role due to the presence of coherent structures in the near-wall region. Another avenue of improvement is to consider more realistic particles, such as non-spherical objects (which have been seldomly studied in the context of particle resuspension \cite{brambilla2017adhesion, olivares2024aerodynamic}).



\appendix

\section{Selection of thresholds for particle states}
\label{app:thresholds} 

In this section, we detail the choice behind the two thresholds used to identify particles in their states described in Sect.~\ref{sec:part_dyn:setup} (i.e. moving/resting and attached/detached). This requires a threshold for the fluid velocity at which particles are considered resting (see Sect.~\ref{app:threshold_vel}) and a threshold for the distance at which two neighboring particles are considered to be in contact (see Sect.~\ref{app:threshold_dist}.

\subsection{Threshold on the fluid velocity}
\label{app:threshold_vel} 

\begin{figure}[h]
	\centering
	\includegraphics[width=0.95\linewidth]{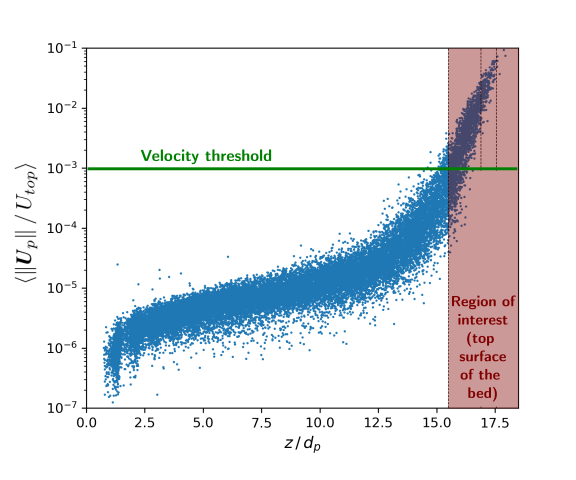}
	\caption{\centering Magnitude of particle velocities normalized by the velocity of the top wall $U_{top} = 0.03$ as a function of the dimensionless elevation $z/d_p$. The proposed threshold allows to select only the particles close to the top as moving ones.}
\label{fig:threshold_vel}
\end{figure}

The first threshold on the fluid velocity has been chosen by analyzing the magnitude of particle velocities $\|\bm{U}_p\|$ (with $\|.\|$ the L2 norm of a vector) as a function of the dimensionless elevation $z/d_p$, which is displayed in Fig.~\ref{fig:threshold_vel}. This confirms that particles that are deep within the sediment bed can have a non-zero velocity (due to local re-arrangement). Yet, in the present study, we are solely interested in the dynamics of particles located close to the top surface of the bed, which is several orders of magnitude larger. For that reason, we have retained a velocity threshold $\epsilon_u = 0.001$ (meaning that particle moving at velocities smaller than 0.1 \% of the top wall are considered as non-moving ones).

\subsection{Threshold on the contact distance}
\label{app:threshold_dist} 

\begin{figure}[h]
	\centering
	\includegraphics[width=0.95\linewidth]{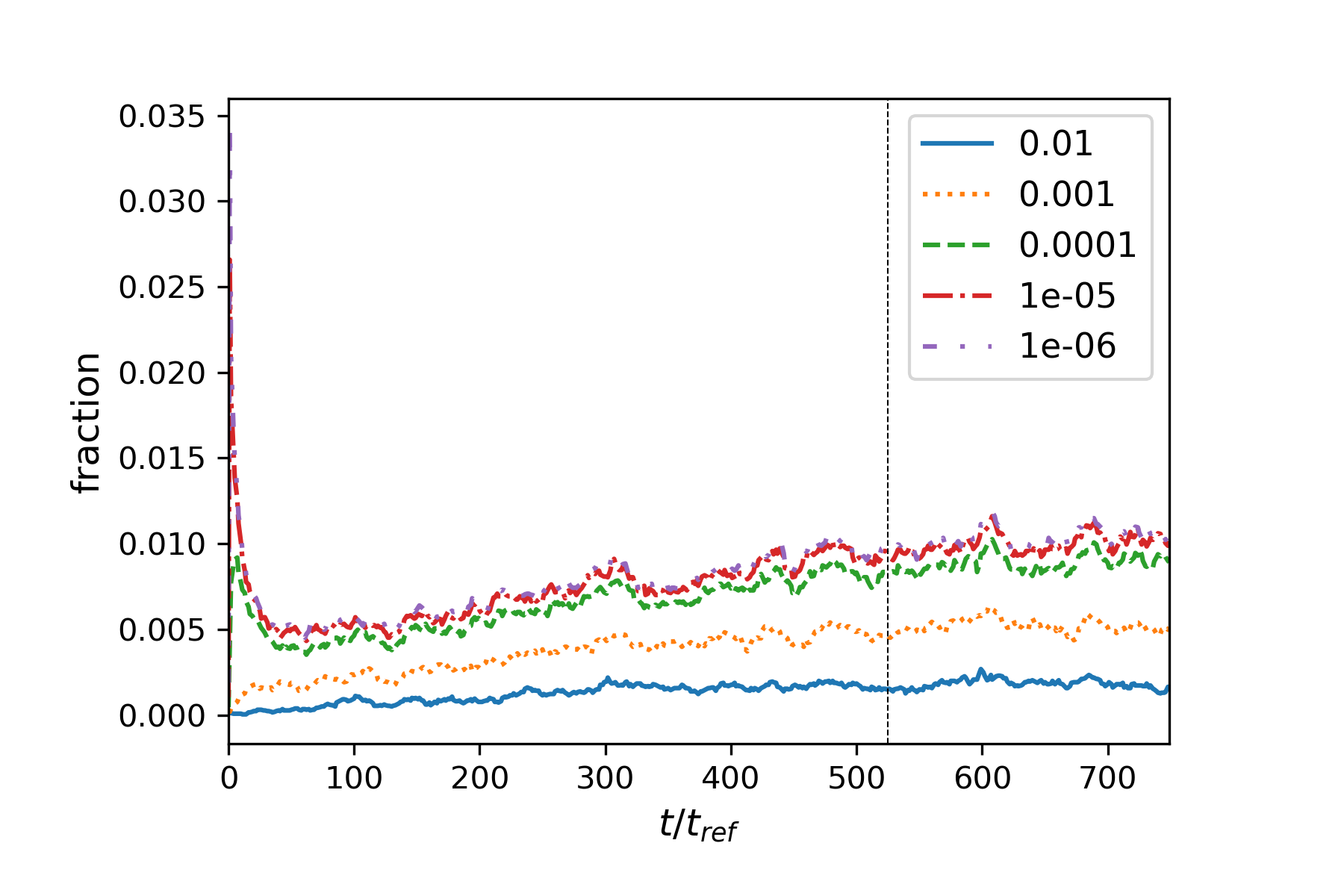}
	\caption{\centering Number of detached particles under different threshold for distance.}
\label{fig:threshold_dist}
\end{figure}

The second threshold on the distance of contact has been chosen by analyzing its effect on the number of particles that are considered attached/detached. For that purpose, we varied the value of the threshold distance in the range $\epsilon_d \in [10^{-2},\, 10^{-6}]$ and we analyzed the results obtained for the fraction of detached particles with the dimensionless time $z/t_{ref}$ displayed in Fig.~\ref{fig:threshold_dist}. It appears that the number of detached particles increases with time until reaching a plateau in the steady-state regime (here for $t/t_{ref}>525$). Yet, the plateau value depends on the value of the threshold considered, except when the threshold distance is smaller than $10^{-4}$. This observation initially convinced us to retain a threshold value $\epsilon_d = 10^{-4}$ since it offers the best compromise between accuracy and efficiency. Nevertheless, it can also be seen that, when using $\epsilon_d < 10^{-4}$, the number of detached particles reaches very high values in the initial stages (here for $t/t_{ref}<20$): this is due to the fact that such small thresholds allow to capture the separation between two particles that are undergoing re-arrangement deep within the sediment bed. Such separations typically last for very short times since particles rapidly reconnect to each other or to other neighbors. Since we are interested here in capturing the detachment of particles on top of the bed (and not extremely quick hop-like motion), we have retained a value $\epsilon_d = 10^{-3}$.

\section{Statistics on particle velocities within the bed}
\label{app:PartVel} 

\begin{figure}[h]
	\centering
	\includegraphics[width=1.0\linewidth]{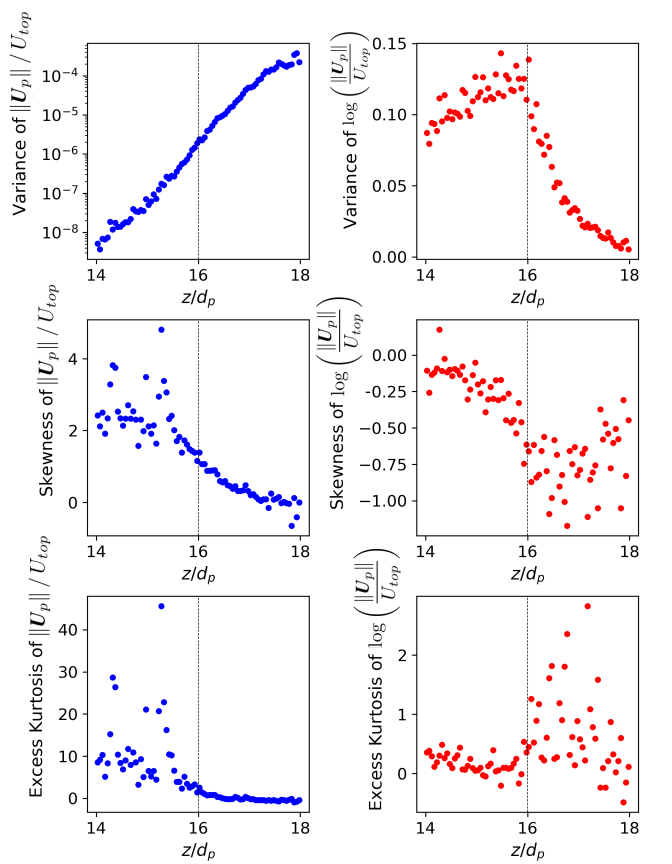}
	\caption{\centering Evolution of the various moments (standard deviation, skewness and kurtosis) as a function of the dimensionless elevation $z/d_p$ for both the magnitude of the particle velocity (left) and its logarithmic value $\log \|\mathbf{U}_p\|$ (right column).}
\label{fig:PartVelMoments}
\end{figure}

In addition to the average particle velocity displayed as a function of the dimensionless elevation in Fig.~\ref{fig:PartVel} and to their distribution in Fig.~\ref{fig:PartVelDistr}, we have computed higher-order moments to further characterize the distribution obtained. This includes the standard deviation of the particle velocity, its skewness and its excess kurtosis. The results are displayed in Fig.~\ref{fig:PartVelMoments}. These plots confirm some of the trends mentioned in Section~\ref{sec:part_dyn:result}: (a) the excess kurtosis and skewness of the particle velocity close to 0 at the top of the sediment bed is in agreement with the observation that the particle velocity distribution resembles a normal distribution, while (b) the log-normal distribution deeper in the bed around $z/d_p\sim14$ is confirmed by the excess kurtosis and skewness close to 0 when focusing on the logarithm of the particle velocity.

\section{Statistics on the surface elevation}
\label{app:elevation}

\begin{figure}[h]
	\centering
	\includegraphics[width=0.99\linewidth]{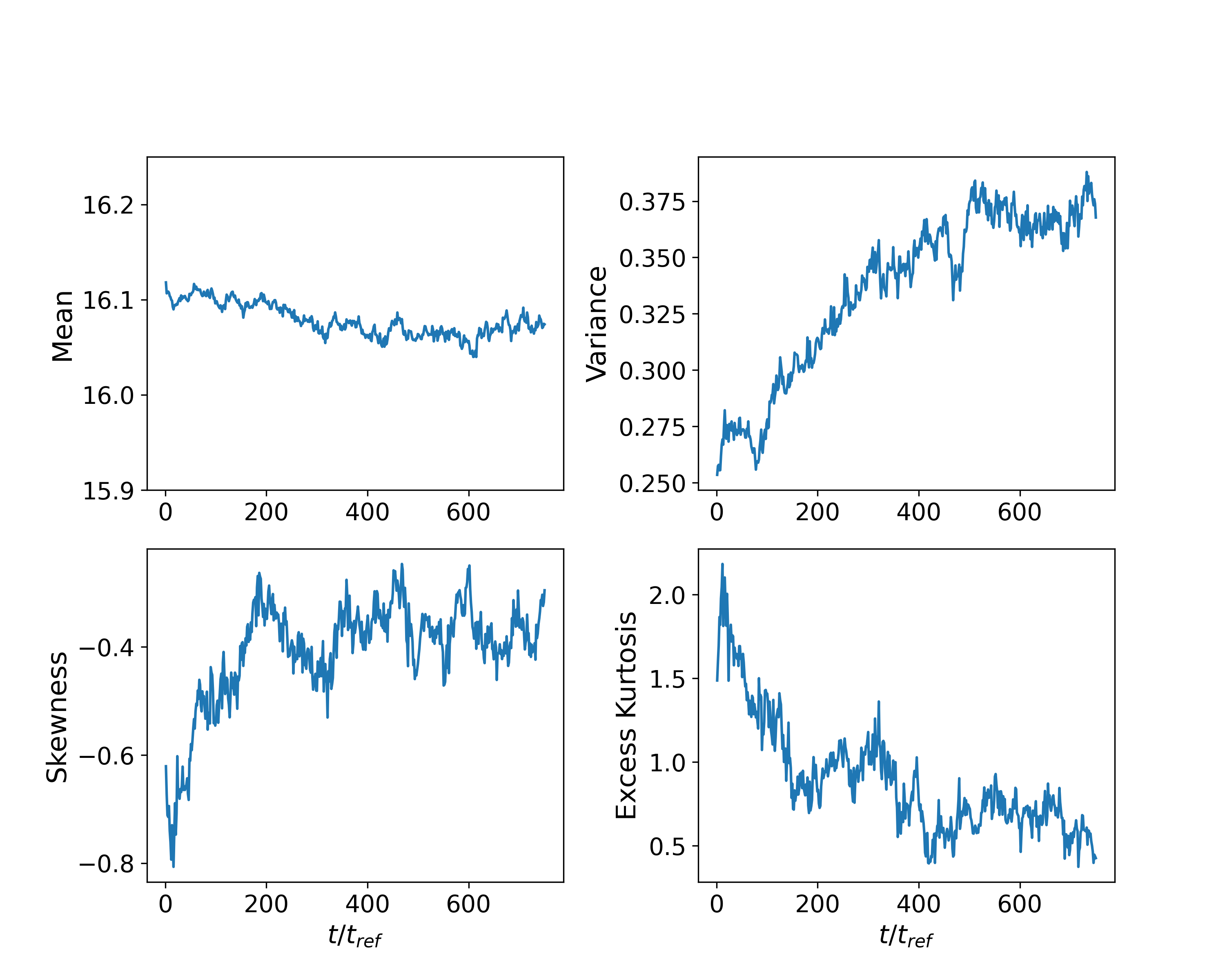}
	\caption{\centering Moments of heights of particle centers on reconstructed surface.}
\label{fig:surface_moments}
\end{figure}

From the reconstructed top surface shown in Fig.~\ref{fig:top_surface}, one can extract statistics on the elevation of the top surface. Figure~\ref{fig:surface_moments} shows the evolution of its average value, its variance, its skewness and its excess kurtosis. It can be seen that these statistics converge to a steady-state regime after 350\,000 iterations, further confirming the establishment of such a steady-state regime. The mean value converges towards $321.4$ while the variance remains relatively large (around $145$), which is indicative that the bed height displays large fluctuations due to the formation of peaks and valleys (as visible in Fig.~\ref{fig:top_surface}). In addition, the values of the skewness is smaller than $-0.5$ while the excess kurtosis reaches values around $0.5$. This tends to indicate that the distribution of heights is comparable to a normal distribution. However, as displayed in Fig.~\ref{fig:pdf_surface_height}, the distribution is indeed close to a normal distribution but it displays some important characteristics that are not captured by a normal distribution such as: a two-peak distribution close to the peak distribution and non-Gaussian tails near the min/max values.

\section*{Authors' contributions} 
The present study was conceived by CH, and later designed through regular exchanges with MB and BV. The study relies on the results of numerical simulations provided by BV and his coworkers. HL has developed the algorithm to analyze the results. The results were analyzed by HL and CH. CH and HL drafted the manuscript. All authors read and approved the manuscript.

\section*{Data access}
The data that support the findings of this study are available from the corresponding author on 
request.


\section*{Acknowledgement}
The authors acknowledge Christoph Rettinger for providing additional data coming from PR-DNS simulations. CH, MB and HL acknowledge Fr{\'e}d{\'e}ric Cazals for useful and constructive discussions on the data analysis using KDTree-Graph approaches. BV gratefully acknowledges the support by the German Research Foundation (DFG) grant no. VO 2413/3-1 and DRESDEN-concept. 

\bibliographystyle{elsarticle-num}
\bibliography{bibliography}

\end{document}